\documentclass[a4,12pt,twosided,reqno]{article}
\usepackage[T1]{fontenc}
\usepackage[utf8]{inputenc}
\usepackage{authblk}
\usepackage{hyperref}
\usepackage{graphicx}
\usepackage{amsmath}
\usepackage{colortbl}
\usepackage{xcolor}
\usepackage{algorithm}
\usepackage{algorithmic}
\usepackage[section]{placeins}

\definecolor{LightCyan}{rgb}{0.80,0.90,1}
\definecolor{Gray}{gray}{0.85}
\definecolor{LightCyan}{rgb}{0.88,1,1}
\definecolor{DarkCyan}{rgb}{0.44,0.88,0.88}
\definecolor{LightViolet}{rgb}{0.88,0.88,1}
\definecolor{DarkGreen}{rgb}{0.15,0.70,0.35}

\newcolumntype{a}{>{\columncolor{Gray}}c}
\newcolumntype{b}{>{\columncolor{white}}c}

\begin{document}

\vspace*{0.35in}

\begin{flushleft}
{\Large
\textbf\newline{Parameter estimation tools for cardiovascular flow modeling of fetal circulation}
}
\newline
\\
Gabriella Bretti \textsuperscript{1,*},
Roberto Natalini\textsuperscript{1},
Annalisa Pascarella\textsuperscript{1},
Giancarlo Pennati\textsuperscript{2},
Daniele Peri\textsuperscript{1},
Giuseppe Pontrelli\textsuperscript{1},
\\
\bigskip
{1} Istituto per le Applicazioni del Calcolo -- CNR,
Via dei Taurini 19 -- 00185 Rome, Italy
\\
{2} Laboratory of Biological Structure Mechanics, Department of
Chemistry, Materials and Chemical Engineering “Giulio Natta”, Politecnico di Milano,  Piazza Leonardo da Vinci, 32 -- 20133 Milano, Italy
\\
\bigskip
* gabriella.bretti@cnr.it

\end{flushleft}

\section*{Abstract}
Usually, clinicians assess the correct hemodynamic behavior and fetal well-being during the gestational age thanks to their professional expertise,
with the support of some indices defined for Doppler fetal waveforms. Although this approach has demonstrated to be satisfactory in the most of the cases, it can be largely improved with the aid of more advanced techniques, i.e. numerical analysis and simulation. Another key aspect limiting the analysis is that clinicians rely on a limited number of Doppler waveforms observed during the clinical examination. Moreover, the use of simple velocimetric indicators for deriving possible malfunctions of the fetal cardiovascular system can be misleading, being the fetal assessment based on a mere statistical analysis (comparison with physiological ranges), without any deep physio-pathological interpretations of the observed hemodynamic changes. The use of a lumped mathematical model, properly describing the entire fetal cardiovascular system, would be absolutely helpful in this context: by targeting physiological model parameters on the clinical reliefs, we could gain deep insights of the full system. The calibration of model parameters may also help in  formulating patient-specific early diagnosis of fetal pathologies.
In the present work, we develop a robust parameter estimation algorithm based on two different optimization methods using synthetic data. In particular, we deal with the inverse problem of recognizing the most significant parameters of a lumped fetal circulation model by using time tracings of fetal blood flows and pressures obtained by the model. This represents a first methodological work for the assessment of the accuracy in the identification of model parameters of an algorithm based on closed-loop mathematical model of fetal circulation and opens the way to the application of the algorithm to clinical data.

\vspace{2pc}
\noindent{\it \bf Keywords}: 
\noindent{Fetal circulatory system, lumped parameter model, Differential algebraic equations, Simulation and numerical modeling, Parameter estimation techniques, Inverse problem.\\}

\noindent{MSC:}
\noindent{65L80; 81T80; 78M50; 70F17}

\section*{Introduction}
The mathematical study of the adult human blood circulation is a quite consolidated subject, since the first work about it dates back in the late nineteenth-century \cite{frank}. On the contrary, the investigation of fetal (i.e. in utero) blood circulation is much more recent, with the modeling of sheep fetal cardiovascular system based on animal studies \cite{huike} and the first modeling studies of Doppler waveforms in the human umbilical placental circulation, about a century later \cite{thom, thom2}.\\
Doppler techniques are a powerful tool to assess fetal blood circulation as they allow for identification of characteristic blood velocity profiles in the fetal arterial and venous tree during gestation. Abnormal velocity waveforms have been associated with adverse perinatal outcome or cardiovascular diseases \cite{Hecher}. A complete understanding of fetal hemodynamics and circulatory patterns is necessary for the correct application and interpretation of Doppler findings \cite{canadilla, scaling} and for their best diagnostic use.\\
It is worth noting that most of the techniques typically adopted for the
postnatal circulation cannot be applied to investigate in utero blood circulation, and ultrasound based approaches, such as Doppler velocimetry and echographic imaging, are the only ones applicable in the routine fetal surveillance and prenatal diagnosis. Indeed, catheter-based measurements are completely unsuitable because they are too invasive for a fetus (they are applied in few situations, as the only diagnostic possibility for specific
pathological cases), thus practically preventing blood pressure measurements.
Furthermore, magnetic resonance imaging, that is readily used for quantification of blood flow in adult circulation is highly compromised in utero by spontaneous fetal motions.  In turn, these limitations hinder the possibility of evaluating important fetal parameters like vascular resistances and compliances or ventricular and atrial elastances. These parameters significantly evolve during gestation and their values can be indicative of a proper or abnormal fetal development and growth.\\
In comparison with the information deducible from the individual Doppler velocimetric tracings that can be only indirectly related to fetal vascular conditions, computational models have the advantage of providing a more global view on hemodynamics and, when applied to patient-specific cases, could allow the quantification of circulatory parameters that are currently not measurable in the fetus.\\
A number of existing models have contributed significantly to our understanding of the fetal circulation, because they allowed the investigation of the influence of various  parameters on the flow pulse waveforms in fetal districts, with reference to a generic blood circulation. 
Different mathematical approaches were adopted to model the whole human fetal cardiovascular system or an individual portion, and they can be classified as zero-dimensional open-loop \cite{canadilla, oluf, pennati3}, zero-dimensional closed-loop \cite{munneke,pennati,yigit}, one-dimensional open-loop models \cite{guett, guiot, raines, vandenwij}, three-dimensional models \cite{chen, pennati4,  pennati5, wilke}.\\
As mentioned above, the study of the fetal blood circulation is quite complicated because of the impossibility in obtaining clinical measurements of blood pressures and flow rates. The correct hemodynamic behavior at the different stages of the gestation are usually assessed referring to some descriptive indices (e.g. pulsatility indices) defined for each Doppler waveform, compared with existing normal standard ranges \cite{ciobanu}.\\
The use of mathematical models of the human fetal circulation may help to improve the understanding of the hemodynamic factors determining the index values and, most importantly, the development of non-invasive 
mathematical-based forecasting tools based for the early diagnosis of fetal pathologies. Indeed, mathematical models predicting Doppler tracing are of interest for clinicians: unusual shapes of velocimetric waveforms may indicate abnormal values of vascular parameters, due to a direct influence of a specific disease (e.g. placental disease) or as a compensatory effect of another disease (e.g. brain-sparing effect during fetal growth retardation). Nevertheless, the diagnostic significances of the various suggested indices is often limited, as the anomalies of Doppler indices cannot be uniquely associated to a clear cause, with the risk that the pathological state in the vascular system is not detected at very early stages of disease development.

According to the specific scale of the phenomenon to be studied, various degrees of simplification at some levels have been proposed. One of them concerns the geometrical dimension of the model. In the current study, we focus onto zero-dimensional models, or lumped parameters models, where different regions of the vascular system are grouped in blocks and connected together, in analogy with the electrical circuits. The number of blocks is related to the desired degree of detail, and, often, the blocks  form a closed loop where both blood circulation and cardiac chambers are modeled with suitable lumped parameters to describe the whole cardiovascular system. \\

The main advantage of closed-loop model approach is the possibility to describe the entire circulatory system and the blood pressure-velocity relationship in each block with a relatively simple and computationally effective model.
Here, in particular, we refer to a closed-loop lumped model of the fetal circulation developed and validated by Pennati et al.  \cite{pennati}, able to predict the values for a large number of Doppler indices in a healthy human fetus. The model, originally set for a fetus at term of the gestation, was then extended to other gestational months \cite{scaling}. \\
Although the clinical informations provided by blood flow Doppler measurements are very useful to assess the status of the examined fetus,
they are still limited to few measuring sites and then provide a quite partial description of the cardiovascular system. The main goal of
the methodology proposed  in \cite{pennati, scaling} is to assist the clinicians with a computational tool able to interpret the
collected data (blood flow Doppler velocities), estimate blood pressures
(always not measurable) and flows in the fetal vascular regions not
examined and, more generally, assess the vascular status of fetus
across gestational ages. Namely, the main strength of using a closed-loop lumped model is the possibility to build what-if scenarios by investigating the impact of vascular modifications or adaptations due to a disease. For instance, it allows the investigation of the hemodynamic changes when some model parameters are modified (e.g. resistances and compliances, as during a peripheral vasodilation or vasoconstriction occurring in intra-uterine growth retardation), without directly imposing any flows or pressures. On the contrary, open-loop circuits, which describe a limited part of blood
circulation (e.g. the arterial tree) see \cite{canadilla}, imply to assume fixed boundary conditions to the model, when instead they should modify due to parameter changes. Hence, if a lumped parameter model of the fetal cardiovascular system is conceived to be significantly applied in fetal diagnosis, the use of a closed-loop circuit is mandatory.

An additional important step towards a clinical use of the model is to obtain
patient-specific parameters able to describe individual fetal circulations.
This implies the capability of identifying the specific values of all the
parameters based on few clinically available information. In fact, a number
of studies suggesting various approaches to estimate patient-specific
lumped parameters can be found into the literature, applied to either open-
\cite{Canuto, DeVault} or closed-loop \cite{Pant, Schiavazzi} models of the blood circulation. Nevertheless,
these models are generally devoted to the investigation of the postnatal
blood circulation (neonatal or adult, where more and complete clinical data
can be collected), with a single study by Garcia-Canadilla et al. that focuses
on the in utero circulation \cite{canadilla2}. Namely, the fetal arterial circulation was
described in an-open loop configuration where two patient-specific bloodflow
inputs (ventricular outflows) and a reference downstream pressure are
imposed as boundary conditions. A constrained nonlinear optimization
algorithm, minimizing the mismatch between computed and measured
blood velocity waveforms in three fetal vessels, was employed to estimate 13
model parameters. The methodology was applied to 37 real cases (22
healthy and 15 growth-retarded fetuses), and a good the matching of the
target functions was reported.
Nevertheless, when the number of unknown parameters is high, and the
available patient-specific data are limited and affected by uncertainties (e.g., variable heart rates can be observed across the various time tracings), a preliminary analysis is suggested to verify the identifiability of the
parameters, before applying the method to real patients \cite{Schiavazzi}.

 For this reason, as a first step, a virtual patient should be considered, where the clinical targets consisting in time tracings of blood flows and pressures are generated through a forward model solution (for the sake of simplicity,
in the following we will indicate these targets as \emph{synthetic data}).
A deep check of the ability of the identification
methodology to properly estimate the vascular and biomechanical model parameters is crucial before the model can be used as a descriptive tool of the fetal circulatory system and as a predictive tool for an early detection of fetal pathologies.
The use of synthetic data rather than clinical ones is motivated by the
necessity of verifying the accuracy of the parameter estimation algorithm.
As a matter of fact, when clinical data were used, even in presence of a satisfactory
superposition between real and simulated data, we could not establish the
correctness of the identified parameters. Indeed, as quite similar
(considering also the measurement uncertainties) velocity tracings could be
obtained using different sets of parameters, the knowledge of the vascular
parameters associated to the flow or pressures curves used as targets is
crucial in the verification phase.

The plan of the paper is as follows.
In Section \ref{model} we describe the forward mathematical model of the fetal circulation we adopted and its numerical approximation. 
Sections \ref{sec:Optimization} and \ref{sec:enkf} are devoted to the
description of the optimization techniques used to the solve the inverse problem of parameter identification, while Section \ref{par:TestProblems} describes the logic and the details of the implemented test problems. In Section \ref{sec:results} the results obtained with the adopted methods are presented and a compared and a detailed discussion is contained in Section \ref{discussion}. Finally, in Section \ref{concl} we comment on the potential of the proposed parameter-estimation algorithm and we present the future developments of our work.

\section{Materials and Methods}\label{sec:methods}
The present section is devoted to the description of:
\begin{itemize}
\item the mathematical model adopted for the simulation of the cardio-vascular
fetal system;
\item the methods used for the estimation of fetal cardiovascular parameters;
\item the definition of the objective function and the selection of the
initial guess.
\end{itemize}

In the current study, we assume the parameters of the fetal model presented in
\cite{pennati} as reference values: some random perturbations of these parameters are produced, with an increasing level of variability, and a blind search is then attempted in order to get back the target values.
The search algorithm is based on two different methods: one is
represented by a combination of global and local search, where a global
optimization algorithm, namely Parameter Space Investigation ({\tt PSI})\cite{Peri, Torn} is applied in combination with the Levenberg-Marquardt ({\tt LM}) \cite{L, M}
algorithm for a refinement of the results; the other one is the Ensemble
Kalman Filter ({\tt EnKF}), originally developed for data assimilation into
dynamical models for numerical weather forecasting \cite{Eversen}. Although {\tt EnKF} represents a self-consistent solution of the identification problem, the  {\tt LM} will be also considered for a further improvements of the results obtained with {\tt EnKF}.

\subsection{The model of fetal circulation \cite{pennati}} \label{model}
The closed-loop lumped parameter model of fetal circulation is the forward model in our problem and it was introduced and studied in \cite{pennati}.
The simulation algorithm, based on the mentioned lumped model, was developed in order to have a simple tool that is able to describe and investigate the physiology of whole human fetal circulation. Such model is able to reproduce all the fetal sites usually monitored by Doppler analysis and it consists of two major parts: the heart compartment, divided in four blocks, i.e. the right and left ventricles and atria and the vascular beds (arteries and veins). The vascular bed is divided into 19 compliant vascular districts connected by a network of rigid pipes: 
\begin{itemize}
\item the upper part of fetal body includes the cerebral (CA and BR) and brachial (UB) circulation; 
\item the lower part includes renal (KID), hepatic (HE), intestinal (IN), lower members (LEG) and umbilical-placental (PLAC) circulations;
\item five blocks for aorta (AA, AO1, AO2, AO3, AO4) and two blocks (PA1, PA2) for pulmonary arteries;
\item a lung block (LUNG) for pulmonary circulation;
\item inferior (IVC) and superior vena cava (SVC) and the umbilical vein (UV).
\end{itemize}

\begin{figure}
\centering
\includegraphics[scale=1.15]{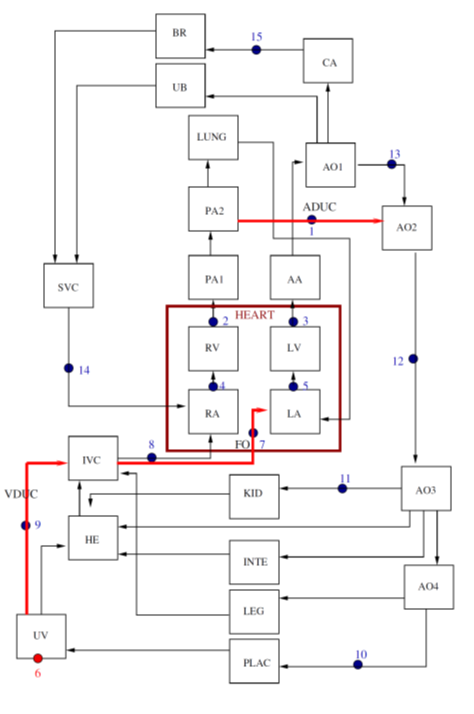}
\caption{Scheme of the lumped parameter model of the fetal cardiovascular system.
Arrows show the normal direction of the flow. Legend: AA: ascending aorta, AO1: aortic
arch, AO2: thoracic descending aorta, AO3: abdominal descending aorta, AO4: femoral bifurcation, BR: brain,
CA: cerebral arteries, HE: liver, INTE: intestinal circulation, IVC: inferior vena cava, KID: kidney, LA:  left atrium, LEG: lower limbs, LUNG: lungs, LV: left ventricle, PA1: main pulmonary artery, PA2: pulmonary
arteries, PLAC: placenta, RA: right atrium, RV: right ventricle, SVC: superior vena cava, UB: upper body, UV: umbilical vein. Shunts (indicated as red bold lines): ADUC: ductus arteriosus; VDUC: ductus venosus; FO: foramen ovale. Blue filled circles: monitored blood flows. Red filled circle: umbilical vein pressure, measured in pathological cases.}\label{fig:distretti}
\end{figure}

See Figure \ref{fig:distretti} for a scheme of the lumped model.

Here we briefly describe the blocks of heart model. Ventricles are
modeled using three components which are related to the
contractile ($U_0$), the elastic ($E_{dia}$ and $E_{sys}$) and the
dissipative viscous ($R_v$) characteristics of the myocardium.
The instantaneous pressure $p(t)$ of the ventricular
model depends on the instantaneous ventricular volume
$V(t)$ and on its derivative $dV(t)/dt$, with an expression based on
the linearization of the pressure-volume-flow characteristics
of a ventricle model including passive and active
myocardial properties, see \cite{pennati} for more details.
Atria are described by a compliance $C_a$ connected
to a time-varying term $U_a(t)$ accounting for the atrial contraction. 
For the four (mitral, tricuspidal, pulmonary, aortic) valves, the instantaneous pressure drop across each valve is computed as the sum of dissipative $K Q(t))^2$ and
inertial $L dQ(t)/dt$ terms, where $Q(t)$ is the instantaneous volumetric flow rate, and $K$ and $L$ are valvular dissipative and inertial flow coefficients. Moreover, the valve flow is one-directional.\\
Each block of the vascular bed is described by
means of a constant compliance $C$, expressing the elastic properties of the vessels. Using the
mass conservation law, the instantaneous
local pressure $p(t)$ is related to the volumetric flow rates
at the inlet and the outlet of the compartment:
$C dp(t)/dt = \sum Q_{in}(t) - \sum Q_{out}(t)$, where $\sum$ indicates
the sum of the flow rates since more than one vessel may enter or exit a compartment,
 and the momentum conservation for each interconnection line holds:
\begin{equation}
\Delta p= R Q + K Q^{\beta} + L {d Q \over dt},
\end{equation}
with $R$ the resistance and $L$ the inertance. In particular, $R Q(t)$ represents the viscous losses along the vessels, $L dQ(t)/dt$ the inertia of the flow, taken into account only for the large arteries close to the heart; the additional term $K Q^{\beta}$, with $\beta \in R^+$, takes into account the local hemodynamics for the fetal shunts i.e., the ductus arteriosus DA, the ductus venosus DV, and the foramen ovale FO.

However, for the sake of brevity, we report in detail the full algebraic-differential set of equations of the model only in the Appendix.\\

By looking at the hydraulic scheme of Figure \ref{fig:distretti}, some interesting peculiarities of the fetal cardiovascular system can be figured out. Indeed, compared to the adult blood flow system which consists of a number of vascular parts arranged as a perfect serial circuit (left heart/systemic arterial circulation/systemic organs/systemic venous circulation/right heart/pulmonary arterial circulation/lungs/pulmonary venous circulation/left heart), in the fetal circulation three important vascular shunts exist which create a hybrid serial-parallel system.
Namely, the shunts are: i) the foramen ovale which directly connects the
systemic venous system with the left heart, ii) the ductus arterious
which directly connects the pulmonary with the systemic circulation and
iii) the ductus venosus which directly links the umbilical vein with
the heart.

Hence, in addition to the normal interconnections between adjacent vascular compartments, these shunts makes the model highly sensitive
even to small perturbations of parameters occurring in quite far blocks,
increasing the complexity of the identification process.

\subsubsection{Numerical solutions of the forward model}

The differential-algebraic equation system describing the model is solved by the explicit Euler method (first order) (together with the Collatz method of second order for the {\tt EnKF}) to integrate the ODE system, see \cite{butcher} for a description of the numerical methods for ODEs, while the algebraic equations are solved using as known data the solutions of the ODE system obtained at the previous time step. The simulation, with an integration step of $\Delta t=4\times 10^{-4} \ s$ and with a fixed set of parameters, is carried out in Matlab$^{\textcircled{c}}$ until the regime configuration is reached (20 cardiac cycles), in about 2 seconds on an Intel(R) Core(TM) i7-3630 QM CPU 2.4 GHz. 

  We remark that we decided not to use the ODE Matlab packages to solve the ODE system since in that case the integration step $\Delta t$ would be variable across the time of the simulation. However, the {\tt EnKF} method used for parameter estimation need to have a fixed integration step in the time window.
  Moreover, to get the code simpler, we implemented explicit Euler's method in the simulation algorithm. Due to the stiffness of the problem, we applied a small time step $\Delta t$ so that to avoid numerical instability of the solutions.  

\subsection{Definition of the objective function}\label{sec:obj}

For our optimization problem the objective function is represented by the distance between the target curves and the curves obtained by using the trial set of parameters. 
In order to carry out our methodological study we need a convergent and stable value of the simulation. After preliminary tests, we observed that the simulation achieves stability and periodicity after about 20 cardiac periods. Then, in order to compare target curves with those obtained by a trial set of parameters, we compare the curve profiles at the 20-th simulated period. The distance between two curves is measured by two different approaches:

\begin{itemize}
\item {\bf Direct measure of the distance - $L^2$:} the more intuitive approach is represented by the sum of squares of the difference between trial and target solutions. Since both the data are produced using the same implementation of the mathematical model, using the same time discretization, we can compare easily the samples, one by one, over a single cardiac period. All the regions where the local solution is naturally zero have been excluded from the analysis, and only one tenth of the points are compared in order to reduce redundancies. We will refer hereafter to this objective function as $L^2$.
\item {\bf Indirect measure of the distance - Analysis in the frequency domain - FFT:} another possible approach is represented by frequency spectrum analysis of the two curves to be compared. Since the observed phenomena are periodic, we can compute the frequency content of the last computed period of the signal by a Fourier analysis, and then compare the amplitude of the first 5 components. We will refer hereafter to this objective function as FFT.
\end{itemize}

The use of different metrics could produce an easier identification of the basin of attraction of the global minimum of the function, hopefully reducing also the multimodality of the problem: for this reason, we tried two different approaches, in order to verify this eventuality. 

Actually, we found the two metrics to be almost equivalent for our purposes, as shown in Section \ref{sec:results}, with a small preference for the $L^2$: in the following we will use the $L^2$ approach, being simpler to compute than FFT.

\subsection{Selection of initial guess for identification using our synthetic data}\label{par:init}

Both {\tt PSI} and {\tt EnKF} methods require an initial guess of the parameters to be estimated. Since the choice of the starting point may have an influence in the final parameters estimation, 10 different sets of initial guess are selected. The random variation of each parameter with respect to the target values is fixed inside a prescribed range (30 $\div$ 40\%): this corresponds to the variability in the curve profiles observed in clinical measurements on healthy fetuses. 

The forward lumped model is solved for all the perturbed sets of parameters, and the pulsatility index (PI) is computed in some crucial sites, see Table \ref{tab:PI}. In more detail, in order to describe the flow velocity in the arterial vessels, the pulsatility index $PI=(V_s-V_d)/V_{mean}$ is calculated using the maximum systolic $V_s$ and minimum diastolic $V_d$ velocities and the mean value of the velocity in the cardiac cycle ($V_{mean}$). The PI index is typically computed by using the flow speed because it represents, in practice, the only direct measure available, see \cite{Arduini}. However, from the fluid dynamic standpoint, we have a conservation
law for the mass flow, and not for the speed. The mass flow is linked to the speed by the section area of the blood vessel:
if the section area can be considered as a constant, the formulation of the PI in terms of mass flow and speed are numerically identical. Since our formulation provides the mass flow, we applied the same formula using the mass flow ($Q$) instead of velocity values, thus we have $PI=(Q_s-Q_d)/Q_{mean}$. 

In Table \ref{tab:PI} the average values of the PI of all 15 equations (including test $\mathcal{B}$ and $\mathcal{C}$) for the 10 sets of initial guess are reported together with their normalized variance, which results to be greater than 50\%. This corresponds, approximately, to consider in the optimization procedure for the identification of the target model parameters in \cite{pennati} (referring to the average, i.e. the 50-th percentile), an initial guess value at $5$-th or $95$-th percentile of the normal range. Then we can conclude that the targets are far enough from each set of initial guess both in the parameter space and in the solution space, thus validating our methodology.  

\begin{table}
\begin{center}
\begin{tabular}{l||l|l|l} \hline
Target time tracings &  E[X] & $\sigma$ & 100*$\sigma$/E[X] \\ \hline
Q$_{ADUC}$ (N. 1)  &  32.1 &   22.4   &   69.8 \\
Q$_{AA}$   (N. 2)  &  12.8 &    7.6   &   59.5 \\
Q$_{APOL}$ (N. 3)  &  22.1 &   18.2   &   82.4 \\
Q$_{TRIC}$ (N. 4)  &  10.6 &    8.5   &   80.0 \\
Q$_{MITR}$ (N. 5)  &   8.6 &    8.8   &  102.9 \\
p$_{UV}$   (N. 6)  &  59.4 &   75.7   &  127.3 \\
Q$_{FO}$   (N. 7)  &  99.2 &   75.8   &   76.4 \\
Q$_{IVC}$  (N. 8)  &  53.4 &   31.8   &   59.5 \\ 
Q$_{VDUC}$ (N. 9)  &  45.6 &   33.3   &   73.0 \\ 
Q$_{AOM}$  (N. 10) &  23.0 &   26.4   &  115.0 \\
Q$_{KID}$  (N. 11) &  41.3 &   26.6   &   64.4 \\
Q$_{AODT}$ (N. 12) &  20.0 &   11.7   &   58.4 \\
Q$_{ARCO}$ (N. 13) &  29.0 &   26.0   &   89.9 \\
Q$_{SVC}$  (N. 14) &  49.0 &   44.1   &   90.1 \\
Q$_{MCA}$  (N. 15) &  23.6 &   18.2   &   77.2 \\
 \hline
\end{tabular}
\vspace{0.5cm}
\caption{Percentage variation of the PI values for the 10 sets of initial guess with respect to the corresponding value for the target point. A percentage variation of the variance larger than 50\% is indicating the good degree of coverage. Note that the numbering of the target measurements corresponds to the number of the clinical site reported in Fig. \ref{fig:distretti}.}\label{tab:PI}
\end{center}
\end{table}

\subsection{{\em Black box} optimization}\label{sec:Optimization}

As mentioned above, our approach to calibrate the model
parameters is to consider the mathematical model of the fetal cardiovascular
system as a {\em black box}: some outputs are produced once a set of parameters is given as input, no matter about how they are obtained.
The black box algorithm is characterized by the following features:
\begin{itemize}
\item outputs are represented by curve profiles for the different elements of the cardiovascular system; 
\item for a given set of target curves, the objective function (to be minimized) is defined as a measure of the distance between the target curve profiles and
those actually provided by the model;
\item using this objective function a mathematical programming problem, whose
solution can be obtained by an appropriate optimization algorithm of the class of the non-linear constrained optimization, is defined.
\end{itemize}

Note that no further information are exploited: the output is not linked with any particular feature of the different parameters, neither special considerations are applied, i.e. the dependence or the connection between two different compartments and their local data,
or the possibility of the identification of a parameter from a restricted number of outputs. An example is reported in
\cite{goulet}, where a number of different optimization algorithms, all included in the {\tt Matlab} toolbox, are shown and applied,
but a critical review is here missing. A more complete analysis of the potentialities of the application of optimization algorithms
for the inverse-problem solution are discussed in \cite{Schiavazzi}, where different original algorithms are presented together with some basic considerations about the nature of the optimization problem, including the necessity of different techniques to be combined together. In fact, we have to consider the specificity of this problem: in particular, the fact that every parameter is acting on a specific region, but it is also influencing all the other regions. 

The cardiovascular system is here considered as a closed circuit, and the variation of a model parameter is reflected (sooner or later) on the full system, with variable strength depending on both the specific weight in the equations where it directly appears and the induced weight on other equations. Thus, the sensitivity of each equation to the variations of parameters is also influenced by those not appearing in it.
Moreover, the influence of some of them can be large in absolute terms on the only compartment where another parameter is directly included: this situation can make it difficult to identify weaker ones.
This is probably the reason why by applying standard minimization algorithms without any particular precaution large and unnatural variations of some parameters are observed: if their local influence is small compared to the others, this is the only
way to produce a sensible effect. \\
For the same reasons, a procedure for the reduction of the number of parameters by linear combination, i.e. using the Principal Components Analysis (PCA), is doomed to failure: due to the excessive difference (sometime larger than one or two orders of magnitude) between the assigned weights, together with a strong dependence of the space location (in the parameter space), we can end up merging together a group of parameters where few of them are largely dominating, so that a small variation of some parameters is also forcing
a very large variation of another, or {\em viceversa}, hindering the optimization procedure based on PCA (or recombination in general).
Such dependence from the position in the parameter space is also requiring a frequent recalibration, thus making the optimization
time even longer. \\
For the same reason, a local search algorithm, based on information of the first or second derivative of the objective function
(gradient vector and/or Hessian matrix), becomes very difficult to apply without limitations or special assessments. 
The line search along the descent direction could be limited to speed up the convergence; however, the great imbalance between the components of the gradient would slow down the process until it becomes unenforceable. \\
A different approach, still involving a classical gradient-based optimization algorithm, is depicted in \cite{Paun}. Here a combination of a Gradient-based algorithm and Monte Carlo Markov Chain (MCMC) is applied to the identification of the parameters of a 1D fluid-dynamic model of the pulmonary circulation. Although the procedure is promising and results are encouraging, we have also to stress how the number of involved parameters is significantly smaller (4 instead of 72) than the present case. \\
For the aforementioned reasons, similarly to \cite{Schiavazzi}, we applied the {\tt PSI} \cite{Torn,Peri} global search method combined with a {\tt LM} local search algorithm, tipically used in applications for curve fitting problems \cite{L,M}, to refine the obtained results, described in the following.

\subsubsection{Global search - Parameter Space Investigation ({\tt PSI})}
The adopted global search algorithm is inspired to a very simple concept depicted in \cite{Torn}: since every point of the design space has, in absence of further informations, the same probability to be the global minimum, the only way to find the global minimum is to sample regularly the design parameter space. Once a first set of measures is available, and one of the point shows the smaller value in the sample set, we can increase the density of the search around the promising point.
For this application, due to the small computational cost of the mathematical model, this strategy is viable, also considering the possibility of parallel computations. We sample regularly a rectangular subset of the design parameter space in a prescribed number of points: all the selected points are evaluated. Such method is known as Improved Distributed
Hypercube Sampling, see \cite{Beach}. Once the best point is selected, a new box (with reduced amplitude) is centered on it. This procedure is repeated as soon as the box dimension is not too small. The parameter space is also bounded, so that each search box is limited to lie into a prescribed region, avoiding extreme shifts.

\begin{figure}[h!]
\includegraphics[scale=0.75]{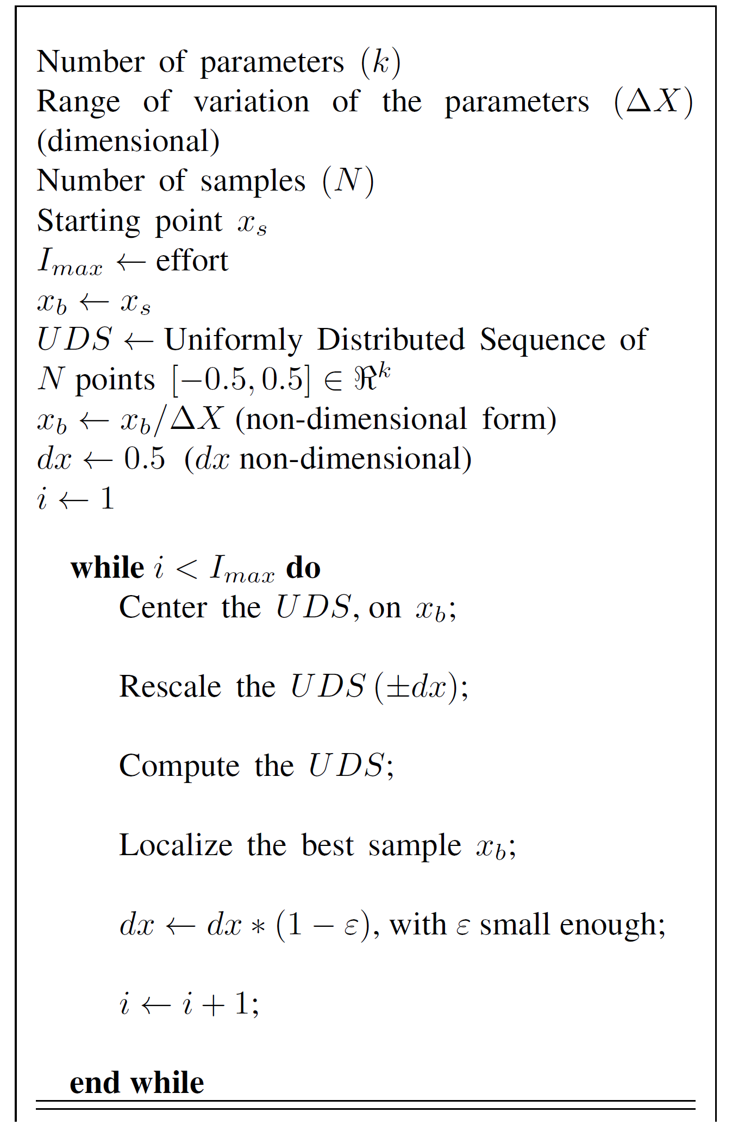}
\caption{Schematization of PSI\, algorithm.}\label{alg:PSI}
\end{figure}

In Fig. \ref{alg:PSI} is reported a pseudocode describing how PSI method works.

We recall here how we are dealing with a complex fetal model, producing 59 outputs (pressure and blood flows of the fetal circulation system). Our implementation of the model depends on 72 parameters\footnote{the atrial and ventricular systolic times and the time displacement between atrial and ventricular activation are here imposed, as they can be considered known on the basis of the heart rate, that is a value clinically measurable. Indeed, given the heart rate, some semi-empirical relationships between those parameters do exist.}, all interconnected to each other.

Due to this very large number of parameters, the spatial density of the investigation cannot be large as soon as the box is not small enough: the absence of clear limits on the variables is a further obstacle to the analysis, so that the reduction of the dimensions of the investigated space represents an outcome of the procedure, and cannot be imposed elsewhere. To give some elements, we can consider how the corner point of a $\Re^k$ space are $2^k$: in case of $\Re^{72}$ they are about $5 \times 10^{21}$, and with this number of points we are only selecting the corner points, without putting a single point in the internal part of the box. Consequently, the global analysis cannot be sufficiently dense to identify the target point with a reasonable precision, and this is the reason why the investigation is performed by a sequence of searches and, as a last step, a local search is needed after the global one. The best we can get from this global analysis is a good initial guess for the following local search.

\subsubsection{Local search - Levenberg-Marquardt algorithm ({\tt LM})} \label{LM_algo}

For the local search procedure, the Levenberg-Marquardt \cite{L,M} implementation available in the FORTRAN library {\tt MinPack} \cite{More} is adopted. This methodology requires an estimate of the objective function and its Jacobian. The original subroutine provides a built-in forward-differences scheme for the Jacobian computation, inappropriate for the use in this context due to the presence of rounding error and numerical noise in general. As a consequence, the required derivatives are approximated by using a 9 points central difference scheme, where a Gaussian filter is applied in order to smooth out the noise: after filtering the data, the extremal points are excluded from the gradient computation. The local search starts from the solution provided by the global search algorithm.

\subsection{Ensemble Kalman Filter algorithm ({\tt EnKF})}\label{sec:enkf}
A completely different approach in parameter identification problem is represented by
sequential filtering methods that take advantage of time varying
measurements. Unlike global methods, where the data are used in a
single batch, sequential methods estimate the unknowns with increasing accuracy
as the measurements arrive over time. When applied to parameters
identification problem, the main idea is to estimate the unknown parameters of a dynamical system from noisy measurements of some component of the state vector.\\
A possible generalization of Kalman Filtering in case of non-linear problem is represented by  the Ensemble Kalman Filter ({\tt EnKF})
algorithm \cite{Eversen}. In \cite{DeVault} {\tt EnKF} was applied  in hemodynamics to estimate tissue/wall material properties or Windkessel parameters. Here, instead, we apply the {\tt EnKF} method to address the parameters identification problem of the lumped model, also in combination with an {\tt LM} local search method to refine the obtained estimates.

Here we briefly recall the  main concepts of the Ensemble Kalman Filter algorithm following the formulation of \cite{Arnold}.

In general, given a system of differential equations:
\begin{equation}
\frac{dx}{dt} = f(t, x, \theta), \ \ x(0) = x_0,
\end{equation}
with $x = x(t) \in \Re^d$ the state vector containing the states of the
system (the $d$ solutions of the system) and $f: \Re \times \Re^d \times \Re^k \rightarrow \Re^d$ a known model function, interpreting the unknown vector parameters $\theta \in \Re^k$ as static state components we can define an augmented state vector:
$$ z(t) = \left[x(t), \theta \right] \in \Re^{d+k}, \ \ \
\frac{dz}{dt} = 
    \left[
         \begin{array}{cc}
         f(t, x, \theta)           \\
         0 \\
        \end{array}
    \right]
= F(t, z). 
$$
The problem of estimating $z(t)$ from noisy measurements $y(t)$ of some
component of the state vector $x(t)$ belongs to the class of filtering problem that can be addressed by using a Bayesian statistical approach opportunely defining an evolution--observation model.

First, we have to characterize an evolution model equation for the state vector
$z(t)$.
Let $y_j$ be the numerical approximation of measurements $y(t)$, $x_j$ the numerical approximation of $x(t_j)$ and $\psi$ a numerical integrator performing the propagation step from time $t_j$ to time $t_{j+1}$, i.e.
\begin{equation}
x_{j+1} = \psi(x_j, \theta) + \nu_{j+1},
\end{equation}
where $\nu_{j+1}$ is the numerical approximation error which we model as a random variable whose probability density is a zero mean Gaussian process. 
If $\hat{\psi}$  is a numerical integrator of higher accuracy, the quantity 
\begin{equation}
\label{error_est}
\gamma_{j+1} = \hat{\psi}(x_j, \theta) - \psi(x_j, \theta)
\end{equation}
represents an estimate of the error in numerical integration and we can assume for $\nu_{j+1}$ a covariance matrix $\Gamma_{j+1}$ based on it, i.e.
\begin{equation}
\label{inn_cov}
V_{j+1} \sim \mathcal{N}(0, \Gamma_{j+1}), \ \ \ \Gamma_{j+1} = \tau^2\mbox{diag}(\gamma^2_{j+1})
\end{equation} 
with $\tau > 1$ a safeguard factor introduced to compensate for the omission of the higher order terms. 

Under the assumption that the estimated error (\ref{error_est}) represents the
standard deviation of an innovation term $v_{j+1}$, the evolution model
equation for the augmented state vector $z(t)$ is given by
\begin{equation}
\label{ev_eq}    
 z_{j+1} = 
    \left[
         \begin{array}{cc}
         \psi(x_j, \theta) \\
         \theta \\
        \end{array}
    \right]
+
    \left[
         \begin{array}{cc}
          v_{j+1}\\
         0 \\
        \end{array}
    \right]
 = \Psi(z_j) + V_{j+1},
\end{equation}
where we assume that the propagation scheme for the static parameter is perfect, i.e. $\theta = constant$.  

Let $y_{j+1} \in \Re^m$ be the observed data at time $t_{j+1}$. Assuming that the measurements consist of direct noisy observations of some components
of the state vector $x(t)$ we obtain the following observation model equation:
\begin{equation}
\label{obs_eq}
y_{j+1} = Bz_{j+1} + w_{j+1}, \ \ \ B= [P \ \ 0_{m \times k}],
\end{equation} 
where $P \in \Re^{m \times d}$ is a projection matrix and the observation noise $w_{j+1} \in \Re^m$ is a zero mean Gaussian process with known covariance matrix,
$w_j \sim \mathcal{N}(0, \mbox{D})$, and independent of $z_{j+1}$.

Given the evolution-observation model defined by equations
(\ref{ev_eq}--\ref{obs_eq}), Bayesian filtering methods allow to sequentially
compute the posterior probability density of the the unknowns $(x_j, \theta)$ in a sequential way:
\begin{equation}
\pi(x_j, \theta | C_j) \rightarrow \pi(x_{j+1}, \theta | C_{j}) \rightarrow \pi(x_{j+1}, \theta | C_{j+1}),
\end{equation}
where $C_j = \{y_1,...,y_j\}$ is the data accumulated up to time $t_j$. In other words, the increasing information is integrated gradually as the data arrive.
Here we use the Ensemble Kalman Filter ({\tt EnKF}) algorithm based on the use of Monte Carlo techniques and ensemble statistics. Indeed, the main idea of {\tt EnFK} is to approximate the covariance matrices of innovation and observation noise with sample covariance matrices computed using the sample mean of a random sample of realizations of the state variable.
\\
The main steps of the implemented {\tt EnKF} algorithm are summarized in (\ref{enkf_appendix}). We also combined the {\tt EnKF} algorithm with the Levenberg-Marquardt ({\tt LM}) local search method described in \ref{LM_algo} to refine the obtained estimates. 

\subsection{Definition of the test problems (synthetic data)}\label{par:TestProblems}

First, we preliminarily tested the algorithm in an ideal situation represented by a target of all 59 time tracings (blood flows and pressures), named set $\mathcal{A}$, to check the applicability of our estimation algorithm. Although this is an ideal and unachievable situation in practice (due to the impossibility of physically obtain pressure measurements in utero and to necessity of minimizing the time for clinical examination), test $\mathcal{A}$ serves to see the applicability of the methodology in presence of the maximum content of information.\\  
Then, two different test cases have been defined with increasing difficulties, in order to check the ability of the algorithms in the identification of model parameter having less and less measurements available, and also provide a quantification of the correctness of the estimate. In particular, we consider as target:
\begin{itemize}
\item 15 time tracings, named set $\mathcal{B}$, i.e. 14 synthetic curves of blood flows profile: descending aorta,inferior and superior vena cava, middle cerebral artery, renal artery, umbilical artery, aortic isthmus, ductus venosus, ductus arteriosus, pulmonary, aortic, mitral and tricuspidal valves) and the quasi-steady pressure profile of umbilical vein;
\item 6 time tracings (a subset of test $\mathcal{B}$), indicated as set $\mathcal{C}$,  i.e. ductus arteriosus, pulmonary, aortic, mitral and tricuspidal valves, plus the quasi-steady pressure profile of the blood pressure of umbilical vein.
\end{itemize}
The motivation for the choice of these sets will be discussed in detail below.\\

The reference curves used as target solution are produced using the model described in \ref{model}. These parameters refer to the final gestational period, when fetal body is about 3 kg. Numerical values can be found in \cite{pennati} (Tables 1 and 2).
In all the cases above, the reference curves are obtained solving numerically the forward lumped model in a time window composed by 20 cardiac cycles as explained more deeply in Section \ref{sec:obj}.\\

In producing the synthetic data, we fix the initial value for each equation of the system, in order to reduce the possible instabilities in the solutions of the forward model due to a change of the parameters; for this reason the same initial values are preserved also during the identification phase.\\
The choice of the compartments of fetal circulation analyzed in the clinical practice may vary significantly, depending on the screening protocol adopted by medical centers. To our knowledge, in the usual clinical practice the number of clinical measurements of healthy fetuses can be around 3-5 curve profiles of blood flows, while in the presence of pathologies the number of measurements can potentially increase until 14. In addition, we may refer
to the pressure values in the umbilical vein taken from the literature, or, in few cases for pathological fetuses, obtained by villocentesis (also known as percutaneous umbilical cord blood sampling).
Following these considerations, in order to reproduce realistic situations, we define two other tests, much more in line with the clinical diagnostics. In these tests, a limited number of equation solutions is adopted. Test $\mathcal{B}$  uses as target measurements the following 15 time tracings: $Q_{AA}, Q_{ADUC}, Q_{APOL}, P_{UV}, Q_{MITR}, Q_{TRIC}, Q_{ARCO},\\ Q_{AODT}, Q_{MCA}, Q_{SVC}, Q_{IVC}, Q_{KID}$, $Q_{AOM}, Q_{VDUC}, Q_{FO}$, while test $\mathcal{C}$ with uses as target solutions 6 time tracings: $Q_{AA}, Q_{ADUC}, Q_{APOL}$, $P_{UV}, Q_{MITR},\\ Q_{TRIC}$, see Figure \ref{fig:distretti} for a schematization of the clinical sites indicated with filled blue dots (blood flows) and red dot (blood pressure). As previously recalled, pressure data cannot be measured: for this reason, we are using a single pressure data (the pressure in the umbilical vein - the only fetal blood vessel that can be catheterized), while all the others are flow data. Selection of the equations for test $\mathcal{B}$ has been driven by the more indicative compartments from the diagnostic standpoint, while test $\mathcal{C}$ is a subcase of test $\mathcal{B}$ composed by a pressure profile and 5 blood flows profiles representative of the circulation in the heart compartment. In the Appendix we indicated by "*" symbol the equations involving the 15 measurements representing the target time tracings, as reported in Table \ref{tab:PI}.\\

\section{Results}\label{sec:results}
The different algorithms presented in Section \ref{sec:methods} are here applied to the test cases $\mathcal{A, B, C}$ to estimate the 72 parameters of the lumped model. We remark that we are using in the target curves, for the 72 unknown parameters, the values reported in Pennati et al. \cite{pennati} (see Section (\ref{par:init}) and the Appendix for more details) as our target. \\
For each test case, starting from the synthetic measurements, we apply {\tt PSI+LM} and {\tt EnKF+LM} by using 10 different set of initial guess as described in \ref{par:init}.

\subsection{Estimates of time tracings}

When all 59 measurements of the different sites of the lumped model are available (test case $\mathcal{A}$), the curve profiles
computed by using the estimated parameters are matching perfectly the target
curves for all 10 runs for both methods.
This shows the effectiveness of the optimization algorithm when the information content is the maximum available. 

\begin{figure}[htb!]
\begin{center}
\includegraphics[width=0.95\textwidth]{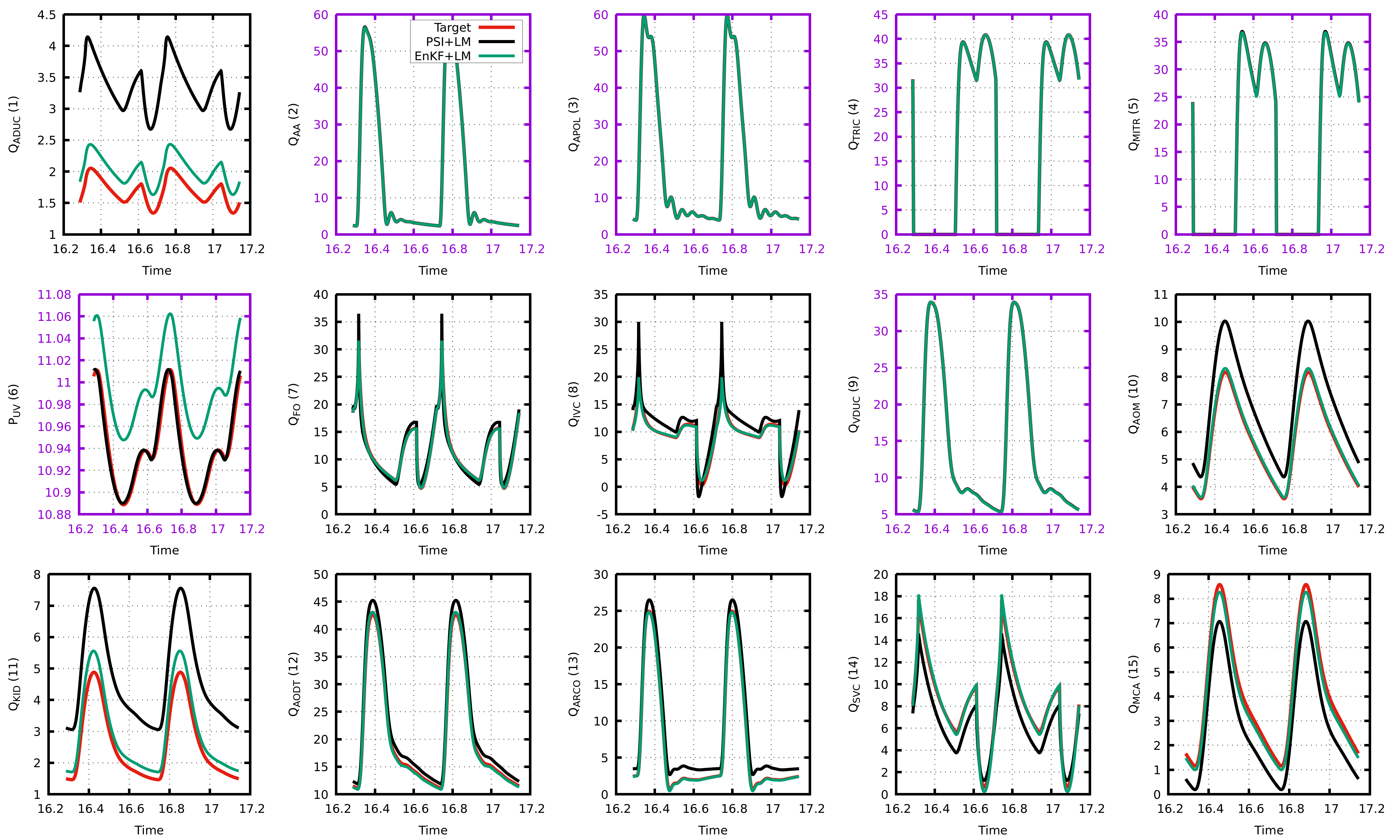}
\caption{Plot of the time tracings for test case $\mathcal{C}$. The boxes of the 6 target time tracings are depicted in purple. The results obtained with {\tt PSI+LM} are reported with black lines, the results obtained with {\tt EnKF+LM} in green and the target ones are in red.}\label{fig:curve6}
        \end{center}
\end{figure}

Test cases $\mathcal{B, C}$ are more significant under a clinical standpoint, since we consider only 15 and 6 measurements, respectively: in particular, while 15 measurements are hardly available in practice, 6 measurements represent a situation pretty typical in the current gestational monitoring. In the case of test $\mathcal{B}$ the target and optimal curves are nearly identical for both methods. More interestingly,
 in Figure \ref{fig:curve6} we depict the last 2 out of 20 cycles of the target
 curves of test cases $\mathcal{C}$ with those obtained by solving the lumped model with the set of the parameters estimated by {\tt EnKF+LM} and {\tt PSI+LM} methods for one set of initial guess. The districts of time tracings in test $\mathcal{B}$ are also included in order to observe the model outcomes for uncontrolled equations. Looking at Fig. \ref{fig:curve6} it is worth noting that the identification of the time tracings is correct even for the districts where no measurements are available. Moreover, the reconstruction of time tracings {\em not included in the group of the target curves} is slightly more accurate with {\tt EnKF+LM}, althought differences with the target are still evident.

\subsection{Estimates of target parameters}\label{estimate}

When all 59 equations are available (test case $\mathcal{A}$), the convergence of the objective functions of the different methods takes place regularly, with negligible differences between estimated and target parameters, thus we do not report in the tables the numerical values of the results obtained in this case.

\begin{figure}[htb!]
\begin{center}
\includegraphics[width=1.05\textwidth]{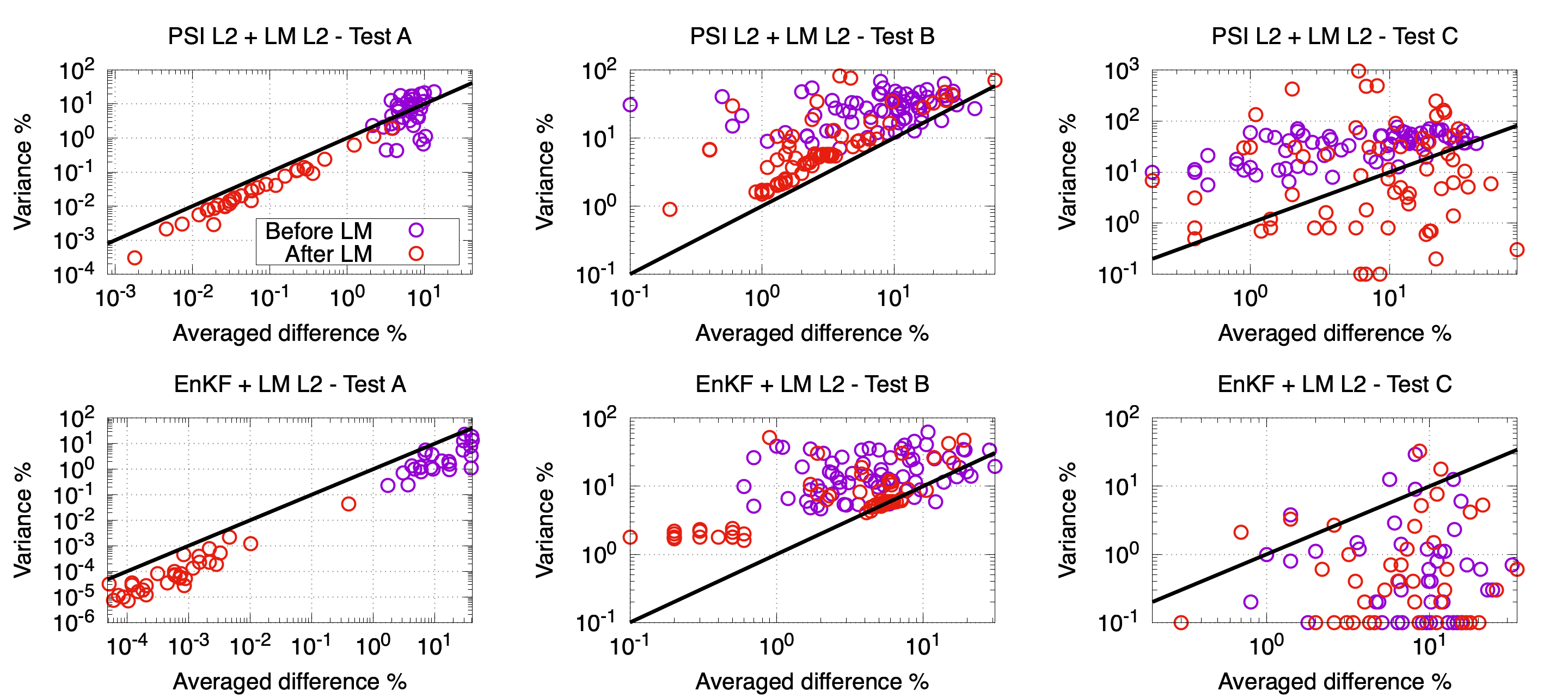}
\caption{Variance and averaged percentage differences of the 72 parameters provided by {\tt PSI+LM} (top) and {\tt EnKF+LM} (bottom). Results for test $\mathcal{A}$ (left), test $\mathcal{B}$ (middle) and test $\mathcal{C}$ (right).}\label{fig:corr}
\end{center}
\end{figure}

In order to analyze the results obtained in the parameter identification we depict in Figure \ref{fig:corr} test cases $\mathcal{A}$, $\mathcal{B}$ and $\mathcal{C}$ on the left, middle and right images, respectively. 

On top, the results from the {\tt PSI} method, before and after the application of the local search are depicted, while {\tt EnKF} is reported on bottom. 
Red dots stand for the results before the local search by {\tt LM}, purple after {\tt LM}. 
The normalized variance (variance divided by the averaged value, here multiplied by 100 in order to compare with percentage difference) of the 10 different results, is reported against the percentage difference between the mean values (over 10 results) of the estimated parameters and the corresponding target values.

The remarkable result revealed by these plots is the substantial correlation
between normalized variance and percentage difference, particularly for test
cases $\mathcal{A}$ and $\mathcal{B}$: this means that, if the variance is low,
the identification of the parameter is also correct, and {\em viceversa}. This
is not a trivial result, since we could have a small variance but a completely
wrong value of the parameter: this would be the case in which all the searches
are pointing to the same incorrect value. Moreover, looking at the results
obtained for test case $\mathcal{B}$ it is evident that the informations brought
by the 15 time tracings are enough to guide the local search method {\tt LM}.
Indeed, after applying it to the results achieved by both {\tt PSI} and {\tt
EnKF} methods, the identification improves and the variance diminishes for all
the 72 parameters, see top pictures in Fig. \ref{fig:corr}. However for test
case $\mathcal{C}$ we do not observe a strong correlation
between normalized variance and percentage difference even after the application of {\tt LM} step.

\begin{figure}[h!]
\begin{center}
\includegraphics[scale=0.1]{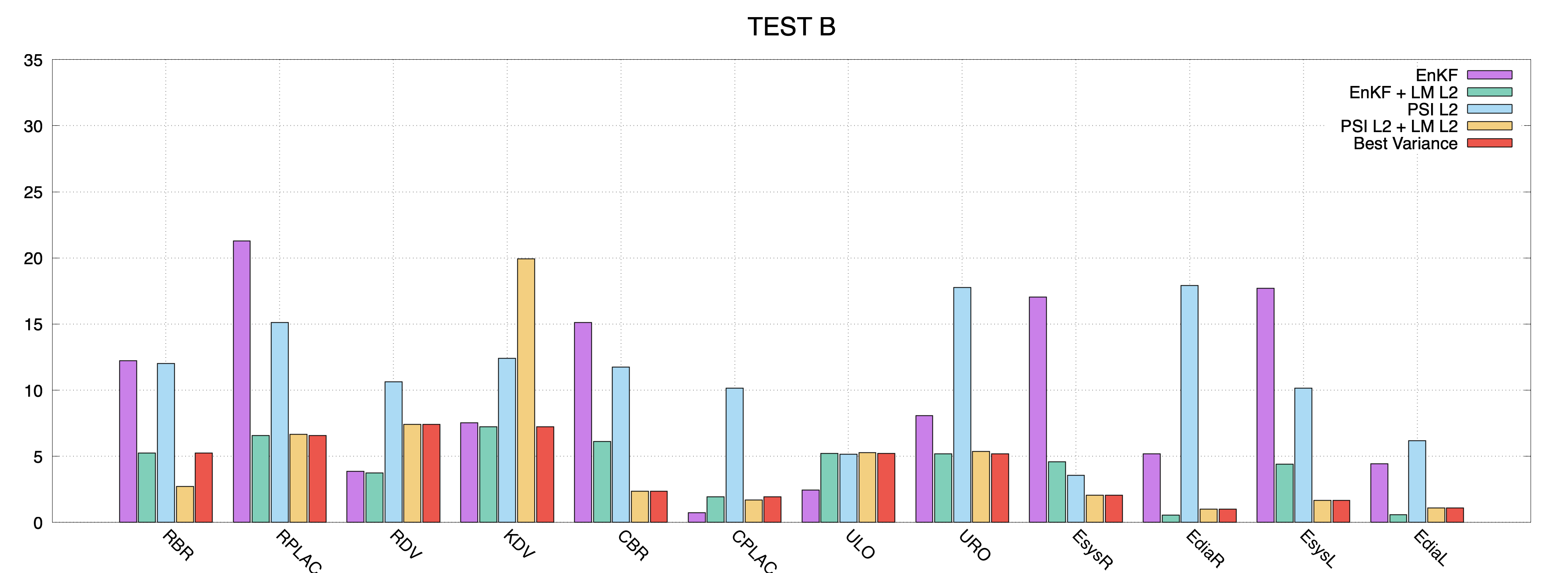}
\caption{Histograms of the $L^2$ percentage differences between target and estimated values obtained with {\tt PSI}, {\tt EnKF} and by {\tt PSI+LM}, {\tt EnKF+LM} using 15 target curves (test case $\mathcal{B}$).}\label{fig:caseB}
\end{center}
\end{figure}

\begin{figure}[h!]
\begin{center}
\includegraphics[scale=0.1]{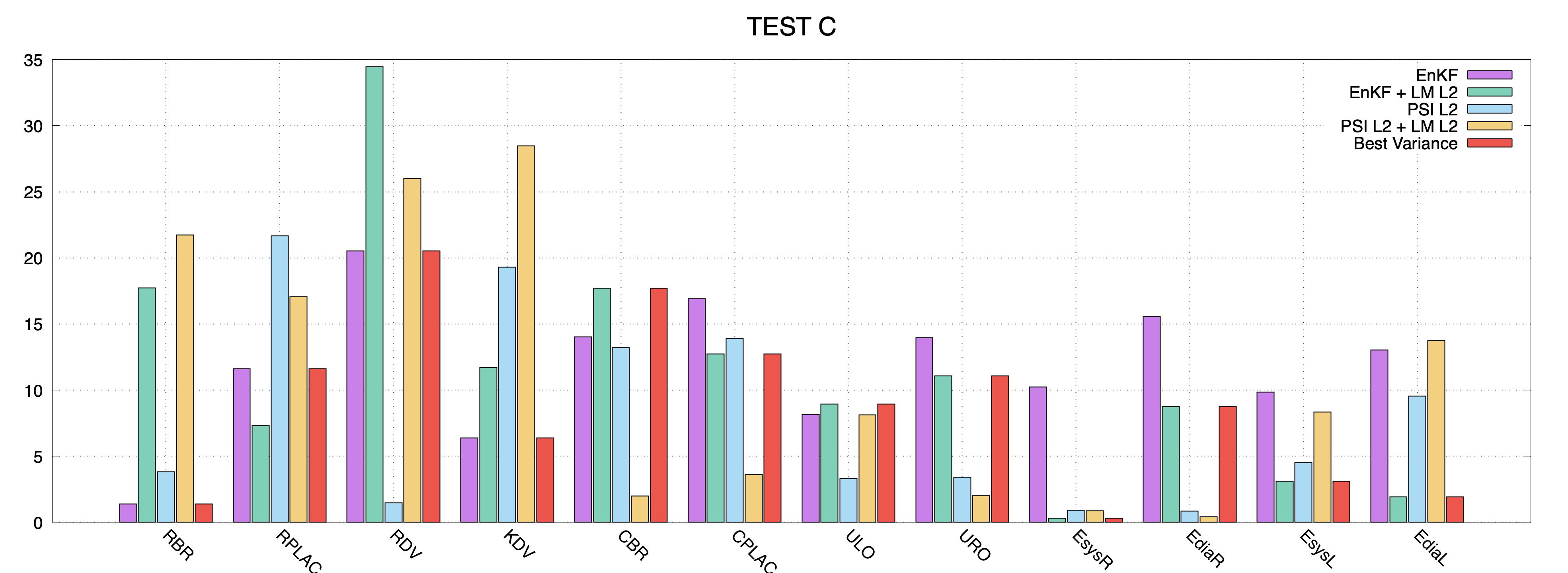}
\caption{Histograms of percentage $L^2$ differences between target and estimated values obtained with {\tt PSI}, {\tt EnKF} and by {\tt PSI+LM}, {\tt EnKF+LM} using 6 target curves (test case $\mathcal{C}$).}\label{fig:caseC}
\end{center}
\end{figure}

Figures \ref{fig:caseB} and \ref{fig:caseC} show histograms of the $L^2$ percentage differences between the mean values of the estimated parameters and the corresponding target values for the 12 most relevant (in the clinical analysis) parameters for the methods: {\tt EnKF} (in purple), {\tt EnKF+LM} (in green), {\tt PSI} (in cyan) and {\tt PSI+LM} (in yellow), for test cases $\mathcal{B}$ and $\mathcal{C}$, respectively.
For test case $\mathcal{B}$ where 15 target time tracings are used,  we can observe that both {\tt EnKF} and {\tt PSI} estimates are improved after applying {\tt LM} local search, see Fig. \ref{fig:caseB} and the related Table \ref{tab:erroriB}. Indeed, both {\tt EnKF+LM} and {\tt PSI+LM} perform well than {\tt EnKF} and {\tt PSI}, respectively, and are able to estimate most of the parameters with a difference around or below 5\% (except for $KDV$, that show a worst estimate, in particular for {\tt PSI+LM}). 

Having only 6 target time tracings (test case $\mathcal{C}$) increases the uncertainty on the estimates, thus causing a loss of accuracy respect to the previous case $\mathcal{B}$. Moreover, having less data also makes more complex the analysis and the discussion of the results achieved, as can be observed in Fig. \ref{fig:caseC} and in the related Table \ref{tab:erroriC}.  In this case, in fact, we can observe that the $L^2$ percentage differences between estimated and target values is not always improved by applying the local search with {\tt LM}. If we look at the results obtained by {\tt EnKF}, the percentage difference stays below 10\% for 4 parameters ($RBR, KDV, ULO, EsysL$) and between 10-20\% for 8 parameters ($RPLAC, RDV, CBR, CPLAC, URO, EsysR, EdiaR, EdiaL$). 
After applying {\tt LM} to {\tt EnKF} results, i.e. {\tt EnKF+LM}, the situation improves significantly for 4 parameters ($EsysR, EdiaR, EsysL, EdiaL$), but, on the other hand, the estimates get worst for 4 of them ($RBR, RDV, KDV, CBR$). 
{\tt PSI} method it is able to produce estimates of 7 parameters with a percentage difference below 5\%, i.e. $RBR, RDV, ULO, URO, EsysR, EdiaR, EsysL$, while the other 4 parameters are estimated with a percentage difference  ranging in $13-22\%$. In the present case with few data available, also for this method the application of the local search with {\tt LM}, i.e. {\tt PSI+LM}, is not so effective, since the estimates are improve significantly only for 2 parameters ($CBR, CPLAC$), and, on the contrary, get much worse for 4 of them.

If we consider that, in a realistic case, we have no information about the target value of the parameter, in 
Fig. \ref{fig:caseB} and \ref{fig:caseC} a last column has been added, reporting the solution, among all the
available ones, with the minimum variance. Particularly for $\mathcal{B}$, this check provides significant information, since the selected solution never overcomes the value of 10\% of percentage difference, and the best solution is picked up correctly in most of the cases. Good indications are also obtained for test $\mathcal{C}$; however, in some cases the best solution is not identified.

 Tables \ref{tab:erroriB} and \ref{tab:erroriC} report the numerical value of the percentage difference between the target value and the identified value of 12 most relevant parameters for the clinical analysis. In the mentioned tables, both average value and its variance, on a sample set of 10 solutions (obtained from 10 different sets of initial guess), are reported. The two different algorithms, {\tt PSI} and {\tt EnKF}, are reported, and the subscript is indicating the adopted metrics, $L^2$ of FFT (as described in section \ref{sec:obj}). The metric does not apply to {\tt EnKF}, that is not dealing with the same objective function as {\tt PSI} and {\tt LM}.\\

In test case $\mathcal{B}$, the best value of the average of the percentage difference between target and identified values is of about 4\% for both {\tt PSI+LM} and {\tt EnKF}, and the best metric is $L^2$. Worst case is around 20\% for {\tt PSI+LM} and 7\% for {\tt EnKF+LM}, then, the maximum error is smaller for {\tt EnKF+LM}.
In conclusion, the results reported in Table \ref{tab:erroriB} validate the choice of $L^2$ as metrics for the objective function.

For test case $\mathcal{C}$ in Table \ref{tab:erroriC} we only report the results obtained by $L^2$ metrics. Note that
the average difference is smaller for {\tt EnFK+LM}, indicating clearly a greater uncertainty for the results provided by {\tt PSI+LM}.

A clearer indication about the effect of a reduced number of target curves is obtained by observing Table \ref{tab:erroriC}, particularly looking at the parameters $ULO$ and $URO$. The variance associated with the identification of
these parameters by {\tt PSI} is significantly larger than for {\tt EnKF}: this is an indication about the larger scattering of the {\tt PSI} data. Once the local search by {\tt LM} is performed, this effect is further amplified
due to the limited number of target curves, and the variance spikes up to 400\% and more. 

Since the local search by {\tt LM} algorithm is the same for both {\tt PSI} and {\tt EnFK}, the different results are only connected with the scattering of the starting points, and a larger dispersion of the solutions occurs due to the multiplicity of the solutions.
This situation is not observed in the results from test $\mathcal{B}$, where the higher number of target curves is probably reducing this multiplicity problem.

The effect of the number of target curves can be also clearly observed in Fig. \ref{fig:corr}. Looking at the results of test
$\mathcal{A}$, after the application of {\tt LM} the points are shifted towards the lower left corner of the graph: this
means that the percentage difference is improved as well as the accuracy of the identification (lower variance). This
situations changes gradually when the number of target curves is decreased: for test $\mathcal{B}$ a similar behavior can still be observed, while for $\mathcal{C}$ this group shift is not so evident.

\begin{table}
{\tiny
\begin{tabular}{l|ab|ab|ab|ab|ab|ab} \hline
                                         &
\multicolumn{2}{|c|}{\bfseries PSI$_{L2}$} &
\multicolumn{2}{|c|}{\bfseries PSI$_{FFT}$} &
\multicolumn{2}{|c|}{\bfseries PSI$_{L2}$+LM$_{L2}$} &
\multicolumn{2}{|c|}{\bfseries PSI$_{FFT}$+LM$_{FFT}$} &
\multicolumn{2}{|c|}{\bfseries EnKF} &
\multicolumn{2}{|c }{\bfseries EnKF+LM$_{L2}$} \\ \hline
\rowcolor{DarkCyan}
Parameter & $\overline{x}$ &  $\sigma$ &
            $\overline{x}$ &  $\sigma$ &
            $\overline{x}$ &  $\sigma$ &
            $\overline{x}$ &  $\sigma$ &
            $\overline{x}$ &  $\sigma$ &
            $\overline{x}$ &  $\sigma$ \\ \hline
RBR     & -12.0 &  34.1 & -10.5 &   4.7 &  -2.7 &   5.2 & -10.1 &   6.5 & -12.2 &   5.9 &  -5.2 &   5.1 \\
RPLAC   & -15.1 &  32.8 & -20.5 &  21.0 &  -6.7 &  17.5 & -19.7 &  12.4 & -21.3 &  14.0 &  -6.6 &   5.9 \\
RDV     &  10.6 &  26.2 &   6.0 &  18.9 &   7.4 &   9.7 &  -1.7 &  18.9 &  -3.9 &  21.6 &  -3.8 &  18.8 \\
KDV     & -12.4 &  44.5 & -14.3 &  42.2 & -19.9 &  32.3 & -26.3 &  40.4 &  -7.5 &  40.4 &   7.2 &  30.5 \\
CBR     & -11.7 &  27.5 & -15.3 &  42.1 &   2.3 &   4.4 & -17.3 &  46.7 & -15.1 &  26.5 &   6.1 &   5.9 \\
CPLAC   &  10.2 &  34.6 &  -2.8 &  44.1 &   1.7 &  10.5 &  -7.7 &  26.2 &   0.7 &  26.2 &   1.9 &   7.6 \\
ULO     &   5.2 &  32.4 & -14.8 &  19.6 &  -5.3 &   8.9 & -17.9 &  12.8 &  -2.4 &  21.9 &  -5.2 &   6.3 \\
URO     & -17.8 &  51.3 & -18.9 &  27.1 &  -5.4 &   9.2 & -21.6 &  17.6 &  -8.1 &  24.1 &  -5.2 &   5.6 \\
EsysR   &   3.6 &  27.7 & -21.2 &  17.9 &  -2.1 &   4.1 & -18.1 &  14.4 & -17.0 &  13.9 &  -4.6 &   5.0 \\
EdiaR   & -17.9 &  39.6 & -19.1 &   8.4 &  -1.0 &   1.6 & -10.1 &   7.9 &  -5.2 &  14.9 &   0.5 &   1.8 \\
EsysL   & -10.1 &  34.5 & -25.4 &  17.1 &  -1.7 &   3.7 & -19.8 &  17.9 & -17.7 &  18.8 &  -4.4 &   4.9 \\
EdiaL   &   6.2 &  27.8 & -19.3 &  14.2 &  -1.1 &   1.6 &  -9.3 &   4.2 &  -4.4 &  11.9 &   0.6 &   2.0 \\ \hline
\rowcolor{LightViolet}
Average &  11.1 &  34.4 &  15.7 &  23.1 &   4.8 &   9.1 &  15.0 &  18.8 &   9.6 &  20.0 &   4.3 &   8.3 \\ \hline
\end{tabular}
\caption{$L^2$ percentage differences in the estimates of model parameters most significant for clinicians. The grey columns indicate the best results obtained. Test case $\mathcal{B}$.}\label{tab:erroriB}
}
\end{table}

\begin{table}
\centering
{\tiny
\begin{tabular}{l|ab|ab|ab|ab} \hline
                                          &
\multicolumn{2}{|c|}{\bfseries PSI$_{L2}$ } &
\multicolumn{2}{|c|}{\bfseries PSI$_{L2}$ + LM$_{L2}$ } &
\multicolumn{2}{|c|}{\bfseries EnKF } &
\multicolumn{2}{|c }{\bfseries EnKF + LM$_{L2}$ } \\ \hline
\rowcolor{DarkCyan}
Parameter & $\overline{x}$ &  $\sigma$ &
            $\overline{x}$ &  $\sigma$ &
            $\overline{x}$ &  $\sigma$ &
            $\overline{x}$ &  $\sigma$ \\ \hline
RBR     &   3.8 &  42.8 &  21.7 & 127.2 &   1.4 &   3.8 &  17.7 &   4.2 \\ 
RPLAC   & -21.7 &  66.1 & -17.1 &  46.2 &  11.6 &   1.1 &   7.3 &   1.2 \\ 
RDV     &   1.5 &  47.9 &  26.0 &  22.6 &  20.5 &   0.6 &  34.5 &   0.6 \\ 
KDV     & -19.3 &  71.2 &  28.5 &   6.2 &   6.4 &   0.1 &  11.7 &   0.2 \\ 
CBR     & -13.2 &  42.1 &  -2.0 &   3.6 &  14.0 &   0.1 &  17.7 &   0.1 \\ 
CPLAC   &  13.9 &  47.1 &   3.6 &  22.8 &  16.9 &   0.7 &  12.8 &   0.6 \\ 
ULO     &   3.3 &  36.3 &   8.1 & 485.2 &   8.2 &   9.1 &   8.9 &   5.2 \\ 
URO     &  -3.4 &  21.2 &   2.0 & 419.7 &  14.0 &  12.7 &  11.1 &   7.7 \\ 
EsysR   &   0.9 &  10.8 &  -0.9 &  30.0 &  10.2 &   0.4 &   0.3 &   0.1 \\ 
EdiaR   &  -0.8 &  14.6 &  -0.4 &   3.1 &  15.6 &   0.1 &   8.8 &   0.0 \\ 
EsysL   &  -4.5 &  26.8 &  -8.4 &  28.8 &   9.9 &   0.6 &   3.1 &   0.1 \\ 
EdiaL   &   9.6 &  33.1 &  13.8 &   3.8 &  13.0 &   0.1 &   1.9 &   0.0 \\ \hline
\rowcolor{LightViolet}
Average &   8.0 &  38.3 &  11.0 &  99.9 &  11.8 &   2.4 &  11.3 &   1.7 \\ \hline
\end{tabular}
\caption{$L^2$ percentage differences and variances in the estimates of model parameters most significant for clinicians. Test case $\mathcal{C}$.}\label{tab:erroriC}
}
\end{table}

\section{Discussion}\label{discussion}
As reported in the previous Section, the differences between estimated parameters and target values are of the same order of magnitude as the expected experimental errors. 
The obtained results satisfy the main requirement for the applicability of our algorithm to real data, represented by Doppler profiles collected by clinicians, if a sufficient number of time tracings is provided.

Regarding the algorithmic point of view, we deduce that both {\tt PSI+LM} and {\tt EnKF+LM} work fine with the largest number of available time tracings. Predictions are still quite accurate in test case $\mathcal{B}$ (15 measurements), with a difference between target parameters and estimated ones around 4-5\% in the average, see Table \ref{tab:erroriB}. The use of only 6 measurements (test case $\mathcal{C}$) makes the identification of model parameters less accurate, with the difference between target parameters and estimated ones around 11\% in the average, but with the presence of estimates with a difference higher than 10\% for some parameters, see Table \ref{tab:erroriC}. 
These results can represent an indication for clinicians that collecting more informations during the Doppler clinical exam of fetuses can contribute to set up an effective mathematical predictive tools of fetal well-being across gestational ages. \\ 
We also derived a possible strategy to assess the quality of parameter estimation also in absence of target parameters for the error computation (in the case of clinically measured curve profiles). This methodology is based on correlation between normalized variance and percentage difference, thus allowing to assess that the identification of the parameter is correct when the variance is low: if the normalized variance is too high, the parameter is not considered for the classification of the fetus.\\
However, looking at the results in the crucial parameters of fetal circulation (regarding heart, ductus venosus, placenta and brain) reported in Tables \ref{tab:erroriB} and \ref{tab:erroriC}, we can conclude that the parameter estimation procedure was successful.
and that the {\tt EnKF+LM} method seems to be more appropriate to provide accurate parameter estimates of crucial model parameters if a limited number of data is available, while {\tt PSI+LM} is substantially equivalent if the number of target curve profiles is higher.
Regarding the computational time of the procedure, the algorithmic structure of {\tt PSI+LM} is largely parallel: {\tt PSI} is currently fully exploiting the {\tt MPI} protocol, while {\tt LM} actually not. On a parallel machine with 36 CPU, the {\tt PSI} analysis takes about 1 hour and 15 minutes. A similar CPU time is required for the
{\tt LM} search. Since we assume 10 different sets of initial guess for the minimal statistical analysis, considering also that the
10 {\tt LM}s can be performed in parallel, the overall time for the analyses is of the order of 12 hours. For this reason, we
cannot apply this methodology {\em on-line}, but the computational time is still compatible with the clinical activity,
considering the approach as a warning for further examinations to be completed the day after. By the way, the measured
curves provided by modern electrocardiograph are ready to use, without the necessity of a digitalization.
On the other side, up to know {\tt EnKF} was implemented without parallelization and
requires less than one hour. The implementation of multiprocessing makes the
algorithm suitable also for a preliminary screening of results.\\
As already mentioned, is of particular relevance to highlight that the proposed approach, requiring only a small amount of data, can provide a picture of all the curve profiles (blood flows and pressures) and the related model parameters (impedances, compliances and resistances) for which we cannot have available measurements, thus enabling the monitoring of the entire fetal circulation system. 
The model, indeed, allows to have informations about locations of the cardiovascular system difficult or even impossible to reach or examine, but it also represents
the prerequisite to attempt, in the next future, a categorization of patients
into pathological and healthy classes according to the statistics obtained from
the estimates of model parameters of a sufficiently large database number of
patients belonging to both classes (pathological and healthy). Moreover, the mathematical algorithm can be seen as a supporting tool to clinicians to simulate what-if scenarios.\\ 
On the other hand, we have also to mention that the manifold interconnections between the vascular compartments make the problem highly sensitive even to small perturbations of model parameters. In addition, it is also important to say that dealing with clinical data, represented by velocimetric blood profiles, will add further difficulties to be considered in the simulation algorithm. In particular, we will need to:
\begin{itemize}
\item add the informations about diameters of blood vessels; 
\item deal with different fetal cardiac frequencies of curve profiles during the Doppler exam to be reconstructed using the same set of model parameters;
\item consider suitable initial conditions to the ODE-algebraic system.
\end{itemize}
Thus, further investigations are required to make the algorithm applicable to real data, with the final goal of constructing a high-fidelity predictive model of fetal circulation.

\section{Conclusions}\label{concl}

Mathematical models for cardiovascular system are largely used to simulate blood flow in arteries and to quantitatively predict dynamical patterns in physiological and pathological conditions. 
The principal feature of the present work has been the development of a simulation algorithm for the parameter estimation of lumped fetal circulation model. Indeed, the problem of identifying the unknown parameters of the model generating a given set of flow and pressure profiles represents a non-trivial inverse problem.
To our knowledge, this work is the first rigorous study on the estimation techniques of model parameters for fetal circulation models. In the existing literature on fetal circulation, see for instance \cite{canadilla2, Canuto},  the errors in the estimation procedure are not quantified, since the algorithms are directly applied to clinical measurements, thus it is also impossible to verify if the reconstructed solution represents the right one among the infinite possible ones. Here, instead, using artificial data as target solutions of the estimation algorithm it was possible to quantify very accurately the errors in the parameter estimation.\\  
To summarize, here we develop a robust calibration algorithm able to perform the accurate fitting of given target curve profiles, even in case of few measurements available, together with a correct identification of target model parameters.

The present work paves the way for future directions of research, as:
\begin{itemize}
\item testing the algorithm synthetic unhealthy patients, in order to check the ability of reproducing more complex situations;
\item applying the calibration procedure on clinical data and building a database with model parameter classification of healthy patients, i.e. normal range for circulation parameters and of pathological patients;
\item using the model as forecasting tool to predict the effects of physiological alterations associated with pharmacological interventions, changes in the environment such as physiological stresses and disease processes;
\item in general, apply the presented methodology to reproduce clinical measurements and of a particular patient (patient-specific) in order to develop a non-invasive forecasting tool to describe the healthy state of the fetal circulatory system across the gestational period.
\end{itemize}

\section*{Credits.}
\noindent G. Bretti: Conceptualization, Methodology, Software, Data curation, Writing - Original Draft, Writing - Review \& Editing, Supervision.\\
\noindent R. Natalini: Conceptualization, Methodology.\\
\noindent A. Pascarella: Conceptualization, Methodology, Software, Data Curation, Formal analysis, Writing - Original Draft, Writing - Review \& Editing, Visualization.\\
\noindent G. Pennati: Conceptualization, Methodology, Software, Data curation, Writing - Original Draft, Writing - Review \& Editing, Supervision.\\
\noindent D. Peri: Conceptualization, Methodology, Software, Data Curation, Validation, Formal analysis, Investigation, Resources, Writing - Original Draft, Writing - Review \& Editing, Visualization.\\
\noindent G. Pontrelli: Conceptualization, Methodology, Writing - Original Draft.\\

All authors have read and agreed to the final version of the manuscript.


\section{Supplementary material}\label{appendix}

\subsection{Full algebraic-differential set of equations of the lumped model \cite{pennati}}
Let us now describe in detail the equations of the model characterized by a differential - algebraic set of 59 equations. We have 29 differential equations:
\begin{align*}
(D1) \quad  & {dp_{AO1} \over dt} = \frac{1}{C_{A01}} (Q_{AA} - Q_{ARCO} - Q_{AO1,2} - Q_{AO1,3}),  \nonumber \\
(D2)^* \quad & {dQ_{AA} \over dt}  = \frac{1}{L_{AA}} (p_{AA}-p_{AO1} - R_{AA} Q_{AA}),\nonumber
 \\
(D3) \quad & {dp_{AO2} \over dt} = \frac{1}{C_{A02}} (Q_{ARCO} + Q_{ADUC}- Q_{AODT}), \nonumber \\ 
(D4) \quad & {d Q_{AO1,2} \over dt}  = \frac{1}{L_{CARO} } (p_{AO1}-p_{CA}-R_{CARO} Q_{AO1,2}),\nonumber \\ 
(D5) \quad & { dp_{AO3} \over dt} = \frac{1}{C_{A03}}(Q_{AODT} - Q_{VRE}- Q_{AO3,2}-Q_{AO3,3}-Q_{AO3,4}), \nonumber \\
(D6) \quad & { dp_{AO4} \over dt} =  \frac{1}{C_{A04} } (Q_{AO3,4} - Q_{AO4,1} - Q_{AOM}),  \nonumber \\
(D7) \quad & { dp_{PA2} \over dt} =  \frac{1}{C_{PA2}} (Q_{APOL} -Q_{PA2,1} - Q_{ADUC}), \nonumber \\
(D8)^* \quad &  {d Q_{APOL} \over dt}   = \frac{1}{L_{PA}}(p_{PA1}-p_{PA2} -R_{PA} Q_{APOL}), \nonumber \\
(D9) \quad & { dp_{LUNG} \over dt} =  \frac{1}{C_{LUNG}} (Q_{PA2,1} -Q_{LUNG}),  \nonumber \\
(D10)^* \quad &  {d Q_{ADUC} \over dt}  = \frac{1}{L_{DA}}(p_{PA2}-p_{AO2}-R_{DA} Q_{ADUC} - K_{DA} Q^2_{ADUC}), \nonumber \\
(D11) \quad & { dp_{BR} \over dt} = \frac{1}{C_{BR}}(Q_{MCA} -Q_{BR}), \nonumber \\
(D12) \quad & { dp_{SVC} \over dt} = \frac{1}{C_{SVC}} (Q_{BR} + Q_{UB}-Q_{SVC}), \nonumber \\
(D13) \quad & { dp_{UB} \over dt} = \frac{1}{C_{UB}} (Q_{AO1,3} - Q_{UB}), \nonumber\\
(D14) \quad & { dp_{IVC} \over dt} =  \frac{1}{C_{IVC}} (Q_{HE} +Q_{VRE} + Q_{LEG} + Q_{VDUC}-Q_{IVC} - Q_{FO}),\nonumber \\
(D15) \quad & { dp_{HE} \over dt} =  \frac{1}{C_{HE}} (Q_{UV,1} +Q_{AO3,2} + Q_{INTE} -Q_{HE}),\nonumber \\
(D16) \quad & { dp_{INTE} \over dt} =  \frac{1}{C_{INTE}}(Q_{AO3,3} - Q_{INTE}), \nonumber \\
(D17) \quad & { dp_{VRE} \over dt} =  \frac{1}{C_{VRE}}(Q_{KID} - Q_{VRE}), \nonumber \\
(D18) \quad & { dp_{PLAC} \over dt} = \frac{1}{C_{PLAC}} (Q_{AOM} - Q_{PLAC}), \nonumber 
\end{align*}

\begin{align*}
(D19) \quad  &{ dp_{LEG} \over dt} = \frac{1}{C_{LEG} } (Q_{AO4,1} - Q_{LEG}), \nonumber \\
(D20)^* \quad  & { dp_{UV} \over dt} = \frac{1}{C_{UV}} (Q_{PLAC} - Q_{UV,1} -Q_{VDUC}),\nonumber \\
(D21) \quad  &{ dp_{CA} \over dt} = \frac{1}{C_{CA} } (Q_{AO1,2} -Q_{MCA}), \nonumber \\
(D22) \quad  & { dp_{AA} \over dt} =  \frac{1}{C_{AA}}(Q_{LV} -Q_{AA}), \nonumber \\
(D23) \quad  & { dp_{PA1} \over dt} = \frac{1}{C_{PA1}} (Q_{RV} -Q_{APOL}), \nonumber \\
(D24) \quad  & {dV_{LV}  \over dt} = Q_{LA}- Q_{LV}, \nonumber \\
(D25) \quad & {dV_{LA} \over dt} = Q_{LUNG} + Q_{FO} - Q_{LA}, \nonumber \\
(D26)^* \quad  & {d Q_{MITR} \over dt}   = \left\{\begin{array}{ll}
0 \ \textrm{ if }  p_{LA}<p_{LV} \textrm{ and } Q_{MITR}<0, \nonumber \\
\frac{1}{L_{Li}} (p_{LA}-p_{LV} -K_{Li} Q^2_{MITR}), \textrm{ otherwise, }
\end{array}\right.\nonumber\\
(D27) \quad & {dV_{RV}  \over dt} = Q_{RA}- Q_{RV}, \nonumber\\ 
(D28) \quad & {dV_{RA} \over dt} = Q_{SVC} + Q_{IVC}- Q_{RA},\nonumber\\
(D29)^* \quad & {d Q_{TRIC} \over dt}  = \left\{\begin{array}{ll}
0 \ \textrm{ if }  p_{RA}<p_{RV} \textrm{ and } Q_{TRIC}<0, \nonumber \\
\frac{1}{L_{Ri}} (p_{RA}-p_{RV}-K_{Ri} Q^2_{TRIC}), \textrm{ otherwise. }
\end{array}\right. 
\end{align*}

The equations above are coupled with the following 30 algebraic equations:
\begin{align*}
(A1)^* \quad & Q_{ARCO} = \frac{1}{R_{ISTHM} }(p_{AO1}-p_{AO2}),\nonumber \\  
(A2)^* \quad & Q_{AODT} = \frac{1}{R_{DTAO} }(p_{AO2}-p_{AO3}),\nonumber \\  
(A3) \quad &  Q_{AO3,4} = \frac{1}{R_{DAAO}}(p_{AO3}-p_{AO4}),\nonumber \\ 
(A4) \quad & Q_{PA2,1}  = \frac{1}{ R_{LUNG} }(p_{PA2}-p_{LUNG}),  \nonumber \\ 
(A5) \quad &  Q_{LUNG} = \frac{1}{ R_{LA}}(p_{LUNG}-p_{LA}), \nonumber \\
(A6)^* \quad &  Q_{MCA}  = \frac{1}{R_{MCA}}(p_{CA}-p_{BR}),  \nonumber \\
(A7) \quad &  Q_{BR}  = \frac{1}{R_{BR}}(p_{BR}-p_{SVC}), \nonumber \\
(A8)^* \quad &  Q_{SVC} = \frac{1}{R_{SVC}}(p_{SVC}-p_{RA}),  \nonumber \\
(A9) \quad &   Q_{AO1,3}  = \frac{1}{R_{UBA}}(p_{AO1}-p_{UB}),  \nonumber \\
(A10) \quad &  Q_{UB} = \frac{1}{R_{UBV}}(p_{UB}-p_{SVC}), \nonumber \\
(A11) \quad &  Q_{HE}  = \frac{1}{ R_{HV}}(p_{HE}-p_{IVC}),  \nonumber \\
(A12)^* \quad &  Q_{IVC} = \frac{1}{R_{IVC}}(p_{IVC}-p_{RA}),  \nonumber 
\end{align*}

\begin{align*}
(A13) \quad &   Q_{UV,1}  = \frac{1}{R_{HA}}(p_{UV}-p_{HE}),  \nonumber \\
(A14) \quad &  Q_{AO3,2} = \frac{1}{R_{ELG}}(p_{AO3}-p_{HE}), \nonumber \\
(A15) \quad &   Q_{AO3,3}  = \frac{1}{R_{MEA}}(p_{AO3}-p_{INTE}),  \nonumber \\
(A16) \quad &  Q_{INTE} = \frac{1}{R_{PORV}}(p_{INTE}-p_{HE}),  \nonumber \\
(A17)^* \quad &   Q_{KID}  = \frac{1}{R_{REA}}(p_{AO3}-p_{VRE}),  \nonumber \\
(A18) \quad &  Q_{VRE} = \frac{1}{R_{REV}}(p_{VRE}-p_{IVC}), \nonumber \\
(A19)^* \quad &  Q_{AOM}  = \frac{1}{ R_{UA}}(p_{AO4}-p_{PLAC}),  \nonumber \\
(A20) \quad &  Q_{PLAC} = \frac{1}{R_{PLAC}}(p_{PLAC}-p_{UV}), \nonumber \\
(A21) \quad  &  Q_{AO4,1}  = \frac{1}{ R_{FA}}(p_{AO4}-p_{LEG}),  \nonumber \\
(A22) \quad  &  Q_{LEG} = \frac{1}{R_{FV}}(p_{LEG}-p_{IVC}), \nonumber \\
(A23)^* \quad  & R_{DV} Q_{VDUC} + K_{DV} Q^2_{VDUC} = p_{UV}-p_{IVC}, \nonumber 
\\
(A24)^*  \quad  & Q_{FO} =\left\{\begin{array}{ll}
\left(\frac{1}{ K_{FO}}(p_{IVC}-p_{LA})\right)^{1/\beta_{FO}}, \textrm{ if } p_{IVC}>p_{LA},\\
0 \ \textrm{ otherwise, }
\end{array}\right.\nonumber \\
 (A25) \quad  & p_{LV} (t)=  U_{LO} A(t) + (E_{diaL} + E_{sysL} A(t)) V_{LV} (t) - Q_{LV} R_{vL},\nonumber \\
(A26) \quad  & K_{LO} Q^2_{LV} +R_{vL} Q_{LV}= p_{LV}-p_{AA},  \textrm{ if }  p_{LV} >p_{AA}, \nonumber \\
& Q_{LV}=0 , \textrm{ otherwise,}\nonumber \\
(A27) \quad  & p_{LA} (t) = U_{aLO} A_a(t) + \frac{1}{C_{aL}} V_{LA},\nonumber \\
 (A28) \quad  &  p_{RV} (t) =  
U_{RV}(t) + E_{RV}(t) V_{RV}(t) - Q_{RV} R_{vR},\nonumber \\
(A29) \quad & K_{RO} Q^2_{RV} +R_{vR} Q_{RV}= p_{RV}-p_{PA1}, \textrm{ if }  p_{RV}-p_{PA1}>0, \nonumber \\
&Q_{RV}=0 , \textrm{ otherwise,}\nonumber \\
(A30) \quad  & p_{RA} (t)=  U_{RA}(t) + \frac{1}{C_{aR}} V_{RA}.\nonumber 
\end{align*}

Note that we indicated by "*" symbol the equations involving the 15 measurements representing the target time tracings reported in Table \ref{tab:PI}.
As mentioned above, the model parameters in the equations of the differential-algebraic system are 72, plus the cardiac times fixed as reported below.
Regarding the unit measure of the model parameters, we have: $U$, isovolumic pressure generator $(mmHg$); $E$, ventricular elastances ($mmHg/ml$); $R_v$, dissipative myocardial resistances ($mmHgs/ml$); $K$ valvular coefficients, ($mmHg s^2/ml^2$), $L$, valvular inertances ($mmHg s^2/ml$); $C$, atrial compliances ($ml/mmHg$).

Note that $A(t)$ and $A_a(t)$ are, respectively, the ventricular and atrial activation functions defined in (\ref{A}) and (\ref{Aa}):
 
\begin{align} \label{A}
A(t)=  \left\{\begin{array}{lcc}
&\left(1/2  \left[1- \cos \left({2 \pi t \over t_s} \right) \right]\right)^\alpha  & 0 \leq t < t_s   \\
& 0    & t_s \leq t < t_c, 
\end{array}\right.
\end{align}
 with $\alpha=0.3$, $t_s$  the systolic period (duration of the miocardic contraction), $t_c$ the cardiac period.
 We also define the normalized sinusoidal activation function $A_a(t)$ for the left and right atria:
\begin{align} \label{Aa}
 A_a(t)=  \left\{\begin{array}{lcc}
\left(1/2  \left[1- \cos \left({2 \pi (t+\tau) \over t_{sa}} \right) \right]\right)^\alpha   & -\tau \leq t < t_{sa} -\tau   \\
 0   & t_{sa} -\tau \leq t < t_c-\tau,
\end{array}\right.
\end{align}
 with $\alpha=0.3$, $t_{sa}$ and $\tau$, respectively,  the duration and the anticipation of the fetal atrial contraction.
Supposing a cardiac frequency of 140 beat/min for a heart model of a normal fetus of 3 kg \cite{pennati}, we assume:
$$
t_s=0.22, t_c=0.43, \tau=0.1, t_{sa}=0.13.
$$

\subsection{EnKF algorithm}\label{enkf_appendix}
We apply the EnKF algorithm to the three different test cases described above with different sets of initial guess values.
We remind that in our case the state system $x$ is given by the $d=59$ equations described in \ref{model} and the parameter vector $\theta$ is the set of $k=72$
parameters of the lumped model.  The main steps of the algorithm are described
in Table \ref{enkf_tab}.  \\
We propagate the system with the one-step Euler method and use the Heun
method to estimate the integration error (\ref{error_est}) and assign the
covariance of the innovation with a fixed time step $\Delta t = 4 \times 10^{-4} s$ and
a sample size of $N=500$. For the initial state $\mathcal{S}_0$, we generate an
initial cloud
$$\theta_0^n = (1 + 0.05 u^n + 0.5 \Delta t \epsilon^n)\theta, \ \ \ n=1, \ldots, N$$
where $u^n \sim \mathcal{U}(0,1), \epsilon^n \sim \mathcal{N}(0,1)$ and $\theta$
of the 10 different sets of initial guess of the three different test cases. To
generate a cloud of initial values for the state system, for each parameter
vector $\theta_0^n$ we propagate the system for 20 cardiac cycles at get the
last state to obtain the initial ensemble $\mathcal{S}_{0|0} = \{ (x_{0|0}^1,
    \theta_0^1), \ldots, (x_{0|0}^N, \theta_0^N)\}$. The observations $y_{j+1}$
    belong to different space based on the test case we are working on, i.e.
    $y_{j+1} \in \Re^m, \ m=59, 15, 6$.

\begin{table*}[h]
	\centering \scriptsize
	\scalebox{0.8}{
		\begin{tabular}{|p{16cm}|}
\hline
	\vspace{.1cm} \textit{Initialization:} Draw the initial combined state and parameter ensemble $ \mathcal{S}_0 = \mathcal{S}_{0|0} = \{ (x_{0|0}^1,
    \theta_0^1), \ldots, (x_{0|0}^N, \theta_0^N)\} $ from the initial prior distribution $\pi(x_0, \theta_0)$.
    Set $j=0$
	\vspace{.2cm} \\
\hline
	\vspace{.1cm} \textit{Prediction step:} Using the current ensemble $S_{j|j}$
    \begin{enumerate}
    	\item Propagates the states using a numerical integrator $\psi$ with specified time step $\Delta t$
    	$$ x_{j+1|j, \psi}^n = \psi(x_{j|j}^n, \theta_j^n, h), \ \ \ n = 1, \ldots, N$$
    	\item Compute the innovation covariance matrix for each ensemble $n$ 
    	$$ C_{j+1}^n = \tau^2\mbox{diag}(\gamma^2_{j+1}), \ \ \gamma_{j+1} = x_{j+1|j, \hat{\psi}}^n - x_{j+1|j, {\psi}}^n$$
    	where $\tau > 1$ is a safeguard factor and $\hat{\psi}$ is a higher order numerical integrator
    	\item Draw $v_{j+1}^n \sim \mathcal{N}(0, C_{j+1}^n) $  for each $n$
    	\item Generate the prediction states via the evolution equation \ref{ev_eq}
	    $$ x_{j+1|j}^n = x_{j+1|j, \psi}^n + v_{j+1}^n, \ \ \ n=1, \ldots, N$$
        \item Combine the predicted state and parameter vectors to form the augmented prediction ensemble
	    $$ \mathcal{S}_{j+1|j} = \{ z_{j+1|j}^1, \ldots, z_{j+1|j}^N \}, \ \ \  z_{j+1|j}^n = (x_{j+1|j}^n, \theta_j^n), \ \ \ n=1, \ldots, N$$
	    \item Compute the prior mean and covariance of the augmented prediction ensemble using the ensemble statistics 
        $$ \bar{z}_{j+1|j} = \frac{1}{N} \sum_{n=1}^N z_{j+1|j}^n
        , \ \ \ \Gamma_{j+1|j} = \frac{1}{N-1} \sum_{n=1}^N (z_{j+1|j}^n -
        \bar{z}_{j+1|j})(z_{j+1|j}^n - \bar{z}_{j+1|j})^T $$
        \item Optionally inflate the prior covariance $\Gamma_{j+1|j} = (1 + \delta)\Gamma_{j+1|j}, \delta > 0$
   \end{enumerate} \\
\hline

\vspace{.1cm} \textit{Observation update:} If an observation $y_{j+1}$ arrives,
    \begin{enumerate}
    \item  Generate the observation ensemble $ \{ y_{j+1}^1, y_{j+1}^2, \ldots,
    y_{j+1}^N\}$ via the formula 
		$$ y_{j+1}^n = y_{j+1} + w_{j+1}^n, \ \ \ n=1, \ldots, N \ \ \ \mbox{where} \
		\ \  w_{j+1}^n \sim \mathcal{N}(0, D)$$
		\item Compute the Kalman Gain
		$$ K_{j+1} = \Gamma_{j+1|j}B^T(B\Gamma_{j+1|j}B^T + D)^{-1} = (\Gamma_{j+1|j}^{-1} + B^T D^{-1} B)^{-1} B^TD^{-1}$$ 
        \item Generate the posterior states and parameters using the updating formula
        $$ z_{j+1|j+1}^n  =  z_{j+1|j}^n + K_{j+1}(y_{j+1}^n - Bz_{j+1|j}^n), \ \ \ n=1,$$
        to obtain the combined posterior ensemble 
        $ \mathcal{S}_{j+1|j+1} = \{ (x_{j+1|j+1}^1, \theta_{j+1}^1), \ldots, (x_{j+1|j+1}^N, \theta_{j+1}^N)\} $
  \end{enumerate}
  Otherwise set $\mathcal{S}_{j+1|j+1} = \mathcal{S}_{j+1|j}$  
\\
\hline
\vspace{.1cm} \textit{Posterior ensemble calculations:}  Compute the posterior state mean and error covariance using the ensemble statistics
	$$ \bar{x}_{j+1|j+1} = \frac{1}{N} \sum_{n=1}^N x_{j+1|j+1}^n, \ \ \ 
   \Gamma_{j+1|j+1} = \frac{1}{N-1} \sum_{n=1}^N (x_{j+1|j+1}^n -
  \bar{x}_{j+1|j+1})(x_{j+1|j+1}^n - \bar{x}_{j+1|j+1})^T $$
  Similarly, compute the parameter mean and covariance using the ensemble statistics
  	$$ \bar{\theta}_{j+1} = \frac{1}{N} \sum_{n=1}^N \theta_{j+1}^n, \ \ \ 
   \Theta_{j+1} = \frac{1}{N-1} \sum_{n=1}^N (\theta_{j+1}^n -
  \bar{\theta}_{j+1})(\theta_{j+1}^n - \bar{\theta}_{j+1})^T $$ \\
\hline
\vspace{.1cm} If $j < T$, set $j = j+1$ and repeat from Step $2$; otherwise, stop.   \\
\hline
	\end{tabular}}
	\caption{Schematization of the EnKF algorithm.}
	\label{enkf_tab}
\end{table*}

\subsection{Sensitivity analysis - completeness of the data}

The large number of parameters adopted in the mathematical model represents a great obstacle for their correct identification. Unfortunately, each parameter is typically characterizing a small part of the whole cardiovascular system: if the area of influence is not included, the measure of the corresponding parameter is obtained only in an indirect manner, and a large variability for that parameter is to be expected. 

\begin{figure}[h!]
\includegraphics[scale=0.25]{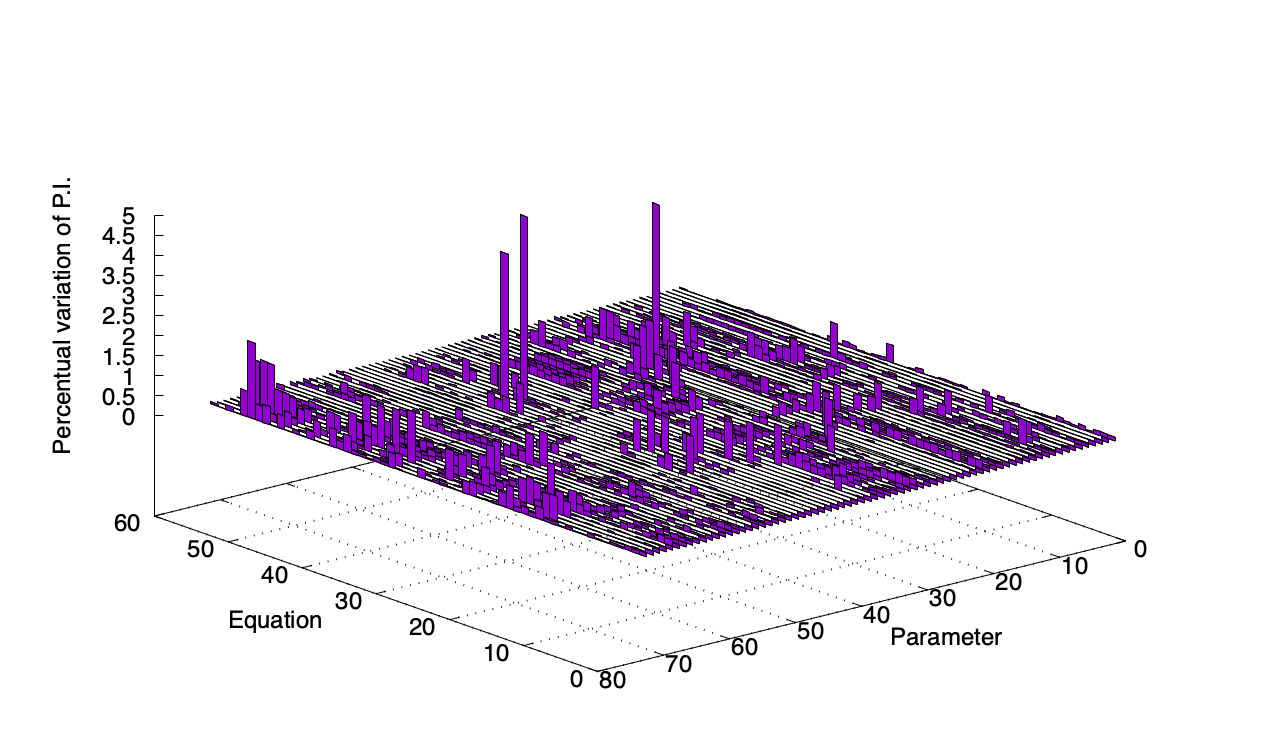}
\caption{Non-dimensional sensitivity of each parameter on each equation. Black is for a very weak influence of a parameter on the corresponding equation.}
\label{fig:sensi}
\end{figure}

In order to give an idea about the real influence of a parameter on a specific equation, a sensitivity analysis has been produced. In this test, each parameter is increased of 1\%, and the influence on the parameter is measured by means of the Pulsatility Index (PI), defined in Section \ref{par:TestProblems}. The results of the sensitivity analysis are reported in Figure \ref{fig:sensi}.

We can see how a very limited number of parameters have a large influence on a subset of the equations. Tipically, the influence of the other parameters is one or two order of magnitude smaller. This situation represent a great obstacle if we try to merge some parameters together, in order to reduce the total number of variables to be considered in the identification problem. On the other side, also the selection of the minimal number of equations necessary for the parameter identification cannot be obtained starting from considerations about the sensitivity. For this reasons, the selection of the equations has been performed on the base of the clinical experience and on the measurements reasonably available.

\end{document}